\documentclass[aps,twocolumn,floats,nofootinbib]{revtex4}
\usepackage{graphics,graphicx,epsfig}
\usepackage{amssymb,color}
\usepackage{epsf,epstopdf,wrapfig}
\usepackage {amsmath}

\newcommand{\beq}{\begin{equation}}
\newcommand{\eeq}{\end{equation}}
\newcommand{\beqn}{\begin{eqnarray}}
\newcommand{\eeqn}{\end{eqnarray}}

\begin{document}

\title{Perspectives on theory at the interface of physics and biology}

\author{William Bialek}

\affiliation{Joseph Henry Laboratories of Physics, and Lewis--Sigler Institute for Integrative Genomics, Princeton University, Princeton NJ 08544\\
Initiative for the Theoretical Sciences, The Graduate Center, City University of New York, 365 Fifth Ave, New York NY 10016}

\begin{abstract}
Theoretical physics is the search for simple and universal mathematical descriptions of the natural world. In contrast, much of modern biology is an exploration of the complexity and diversity of life.   For many, this contrast is prima facie evidence that theory, in the sense that physicists use the word, is impossible in a biological context.  For others, this contrast serves to highlight a grand challenge.   I'm an optimist, and believe (along with many colleagues) that the time is ripe for the emergence of a more unified theoretical physics of biological systems, building on successes in thinking about particular phenomena.  In this essay I try to explain the reasons for my optimism, through a combination of historical and modern examples.  
\end{abstract}

\maketitle

\section{Introduction}

At present, most questions about how things work in biological systems  are answered by experimental exploration. The situation in physics is very different, where theory and experiment are more equal partners.  Almost from the moment that biology and physics became separate  sciences, physicists have hoped that we could reach an understanding of life that parallels our understanding of the inanimate world.  Although there have been several waves of enthusiasm, each with its own successes,  such hopes often have seemed quite fanciful.  Today, as more of the living world becomes susceptible to quantitative experiments, old dreams are being rekindled.  

The increasing body of quantitative data calls out for analysis, sometimes quite desperately, and creates  opportunities to make mathematical models for particular biological systems.  Indeed, the notion of ``modeling'' as part of a modern, quantitative biology is becoming conventional.  But theoretical physics is not a collection of disparate models for particular systems, or a catalogue of special cases.  There is a growing community of theorists who want, as it were, more out of life.  We want  a theoretical physics of biological systems that reaches the level of predictive power that has become the standard in other areas of physics.  We want to reconcile the physicists' desire for concise, unifying theoretical principles  with the obvious complexity and diversity of life.  We want  theories that engage meaningfully with the myriad experimental details of particular systems, yet still  are derivable from principles that transcend these details. 

The existence of a community of optimists does not imply that our optimism is justified.  The goal of this essay is to explain why at least one theorist (me) is optimistic.   I hope to convince you that theory has had important successes, shaping how we think about life today, and that this is true despite a widespread impression to the contrary.  Turning from the past to the present and future, I will argue this is an auspicious time:  theory is having a real impact on experiment, related theoretical ideas are emerging in very different biological contexts, and we can see hints of ideas that have the power to unify and deepen our understanding of diverse phenomena.  What is emerging from our community goes beyond the ``application'' of physics to the problems of biology.   We are asking physicists' questions about the phenomena of life, looking for the kinds of compelling answers that we  expect in the traditional core of physics.

Any effort to justify optimism must be addressed not just to the agnostic, or to the converted, but to the actively skeptical.     Many biologists believe that we just ``don't know enough'' to theorize about the particular systems that they study, and have a hard time pointing to examples where approaches grounded in mathematical thinking have illuminated the workings of these systems. For many physicists, the phenomena of life still  look too messy to be accessible, and they  doubt if there is anything very fundamental to be said, or if digging into the phenomena of life just means sifting through a mass of detail.    My goal in this essay is to respond to these concerns directly.   I hope to convince you that the pessimistic biologists are wrong about the history, and that the pessimistic physicists are wrong about the current state of the field.

In general, it seems best to let the work of the community speak for itself, and provide its own justification for our optimism, rather than making pronouncements about what anyone else should be doing or thinking.  But, in 2014, the Simons Foundation convened the first of what is now an annual series of workshops on {\em Theory in Biology},\footnote{As of this writing, most of the workshop proceedings are  available online, at www.simonsfoundation.org.}  and   I was given the task of providing some perspectives.  This led me to think more explicitly about the grounds for my own optimism, and about the history of theory in our field;  this essay grew out of that short lecture.   It came at the end of a long day and so, perhaps ironically, it was more descriptive than mathematical.\footnote{In addition to the specific references cited, several topics are discussed more fully  (and more mathematically) in Ref \cite{bialek_12}.}

\section{A classical example ($\mathbf\sim$ 1900)}

In the late nineteenth century, continuing through the early 1900s, many of the great figures of classical physics routinely crossed the boundaries between subjects that we now distinguish as physics, chemistry, biology, and even psychology.  In particular, Lord Rayleigh had an   interest in hearing, which he viewed as an extension of his interests in the theory of sound.  In a paper from 1907 entitled ``On our perception of sound direction,''  Rayleigh developed ideas that are probably familiar even if you don't know their origin \cite{rayleigh_07}.

For sounds at  high frequencies,  the wavelength is shorter than the width of your head and thus your head casts the acoustic equivalent of a shadow. So if sound is coming from your right, it is more intense in your right ear than in your left, and  there are plenty of direct experiments to show that this is indeed how you localize high frequency sounds---you pick up on the intensity difference between your two ears.  What Rayleigh understood  was that  if you go to low frequencies this doesn't work: the wavelength becomes longer than the size of your head, and hence your head no longer casts a shadow.\footnote{Although he was responsible for many of the crucial theoretical developments, Rayleigh didn't simply trust  the theory, and actually checked that intensity differences between the two ears of a subject were tiny at low frequencies, despite the fact that we could localize the source of these tones.}  The only remaining clue to the location of the sound source is then the timing or phase difference between your ears.

Of course there's another possibility, which is that you can't actually localize low frequency sounds, so Rayleigh had to check this, and he went on to devise experiments that tested directly whether you could hear the phase or time differences.\footnote{The description of Lady Rayleigh, steadying herself by leaning on a table, with pipes emerging from her ears, is particularly charming.} At this point in history, there was a predominant assumption that we are phase deaf---that we can hear the intensities of the component notes of a sound, but not their phases. But the physics of the situation tells us that if we're going to localize sound at low frequencies then we {\em must} hear phase differences, so there's an immediate qualitative prediction, and this  was confirmed.

Rayleigh phrased his conclusions poetically but accurately:  ``It seems no longer possible to hold that the vibratory character of sound terminates at the outer ends of the nerves along which the communication with the brain is established. On the contrary, the processes in the
nerve must themselves be vibratory, not of course in the gross mechanical sense, but with preservation of the period and retaining the characteristic of phase---a view advocated by
Rutherford, in opposition to Helmholtz, as long ago as 1886.''  In modern language, action potentials in primary auditory neurons must ``phase lock'' to the temporal fine structure of the acoustic waveform.

If we push beyond these qualitative arguments, we find  a surprising quantitative conclusion.  The smallest difference in source direction that we can discriminate, using low frequency tones,  corresponds to a difference in time between our two ears of only a few microseconds---and if you were a barn owl it would only be one microsecond.   This is even more startling since the characteristic time for everything to happen in the nervous system is a millisecond, not a microsecond.  It was more than 50 years before anyone recorded from a neuron that actually implemented these timing comparisons, and it took even longer to demonstrate that precision really is in the microsecond range. 

While one example doesn't make a rule, we can try to identify a strategy at work in this example.   Rayleigh started with  a few facts about biology, and added a few basic physical principles.  Thinking hard about how these connect (or conflict), he arrived at a theory.  This theory made qualitative predictions, and provided  a new framework for quantitative discussion.  This framework in turn yielded startling results, and set in motion a sequence of experiments that played out over many decades.

\section{A more famous example ($\mathbf\sim$ 1950)}

\begin{figure*}
\centerline{\includegraphics[width = 0.9\linewidth]{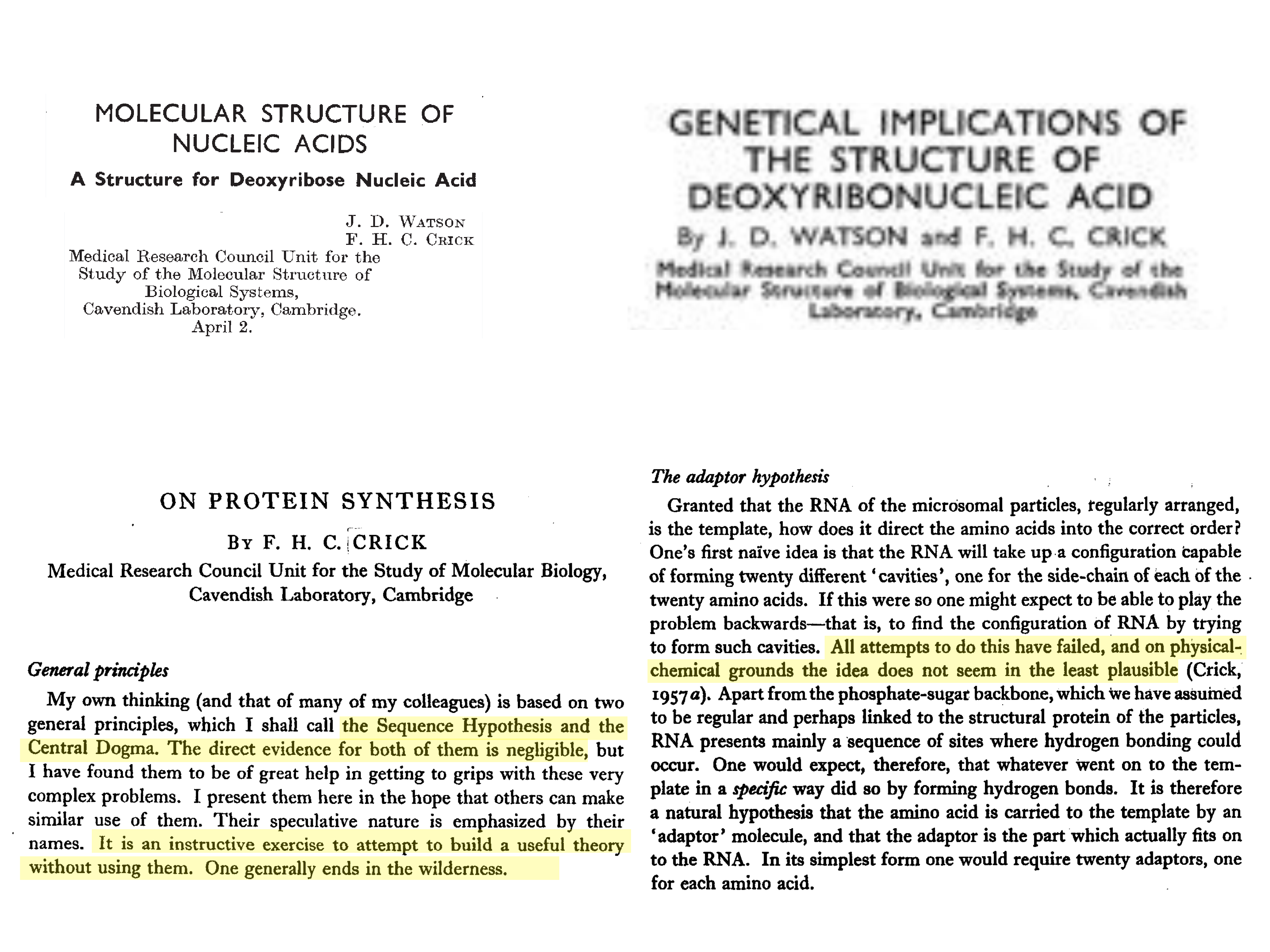}}
\caption{Some classic papers from Watson and Crick (above) \cite{watson+crick_53a,watson+crick_53b}, and from Crick alone (below) \cite{crick_58}.  I have highlighted passages that emphasize the theoretical character of this work, as explained in the text. \label{watson+crick}}
\end{figure*}

Figure \ref{watson+crick} shows another example, one perhaps more familiar to most of you \cite{watson+crick_53a,watson+crick_53b,crick_58}.  None of the papers cited here have original data, and so, by that criterion, these certainly are theoretical papers. I think it's deeper than that.  These papers   follow the pattern that I just suggested to you based on Rayleigh's work: start with a small number of biological facts, add  some basic physical principles,
and mix carefully.  

As you know, Watson and Crick  were trying to build models for the molecular structure of DNA, and in such an effort it is crucial that there are {\em rules} of chemical bonding, not suggestions about chemical bonding. So there are real things, quantitative principles from physics and chemistry, on which one can rely.   And, in trying  to fit all these things together, there appear not to be that many solutions.   You know what happened next.

The first paper by Watson and Crick ends with the cryptic remark about how it has not escaped their attention that the structure they propose has implications \cite{watson+crick_53a}, and in the second paper those implications are worked out \cite{watson+crick_53b}.   The intellectual shockwaves which propagated outward from Refs \cite{watson+crick_53a,watson+crick_53b} are so well known that they don't require a review here, but it's important to look back at what really got said and done in these papers.\footnote{The American philosopher LP (Yogi) Berra is reported to have quipped ``that restaurant is so crowded nobody goes there anymore.'' Perhaps some papers are so famous that nobody reads them anymore.}   Let me note, in particular, the passage ``... any sequence of the pairs of bases can fit into the structure.  It follows that in a long molecule many different permutations are possible, and it therefore seems likely that the precise sequence of the bases is the code which carries the genetical information.  ... one chain is, as it were, the complement of the other, and it is this feature which suggests how the deoxyribonucleic acid might duplicate itself.'' \cite{watson+crick_53b}.

It is crucial to appreciate that these theoretical predictions are {\em not} consequences of experimental observations.  As is well known, parallel to the model building efforts of Watson and Crick in Cambridge, X--ray diffraction experiments on DNA were being done by Franklin, Wilkins, and colleagues  in London  \cite{franklin+gosling_53,wilkins+al_53}.\footnote{The events leading to the initial trio of papers \cite{watson+crick_53a,franklin+gosling_53,wilkins+al_53} are among the most thoroughly studied episodes in the history and sociology of science; for examples see Refs \cite{judson_79,watson_80,maddox_02}.  I still suspect that the essentially theoretical nature of the early work by Watson and Crick has, nonetheless, received less attention than it should.}  Franklin's famous photograph fifty--one, which appears in Ref \cite{franklin+gosling_53}, provided qualitative evidence for a  helical structure, and made it possible to read off the basic dimensions of the helix; these results were quickly clarified in a sequence of papers from Franklin and Gosling \cite{franklin+gosling_53b,franklin+gosling_53c,franklin+gosling_53d}.  But, nearly a decade later, X--ray diffraction data still were not of high enough resolution to ``see'' the pattern of complementary base pairing without relying on models to help interpret the data \cite{langridge+al_60a,langridge+al_60b}.  Langridge et al  provide a very clear discussion of how the data in 1960 were sufficient to test a proposed structure, but not sufficient to determine the structure directly \cite{langridge+al_60b}.     So, in 1953, base pairing was a theory.   And no amount of structural information alone would be sufficient to conclude that ``the sequence of bases is the code which carries genetical information.''  That was a theory too.

The idea that the sequence of bases forms a code defines the problem of deciphering this code, and this attracted attention from many theorists; a highlight from this period is Crick's 1958 paper  \cite{crick_58}, with excerpts  in Fig \ref{watson+crick}.  This is, I think, the paper in which the ideas and phrasing that I have emphasized in the figure appear for the first time. We still use the words ``central dogma,'' but by now  the ``sequence hypothesis''  is so fully internalized that we don't even give it a name. But, as the text states quite explicitly, there was no direct evidence for either proposal. And I invite you to note some of the language that Crick uses, again to emphasize the theoretical character of what was going on:  I tried to build explanations that didn't use these ideas and I couldn't.  

In the decade or so between the proposal of the double helix and the working out of the genetic code, many theorists proposed coding schemes that were quite interesting mathematically, but we now know that none of these proposals was the one chosen by Nature.  Further, by the time the experiments which mapped the code were coming to fruition, nobody doubted that there is a genetic code---that is, the ``sequence hypothesis'' had become obvious, and the problem was to work out what the sequences meant. The combination of these two facts obscures the essentially theoretical foundations of the subject.  As far as I know, all of the experiments that discovered the key features of the genetic code were designed with the theoretical ideas of Ref \cite{crick_58} in mind.

\section{Lessons and problems}

There still are people who ask whether theory will someday, in the distant future, make a contribution to biology.  Thus it is essential to point out that theory  already has made contributions, and big ones at that.  Many of the foundational papers in what we now call molecular biology were unambiguously theoretical papers, and the example of Rayleigh points to a theoretical tradition that reaches  much farther back into the  history of interactions between physics and biology.  But these examples also have problems. 

First, in the case of Watson and Crick, it appears that all the theorizing was in words and not in equations, and so what's written in these papers doesn't look like theory in the sense that we use the term in physics.   I'm not sure that's really fair, because when they went to build a molecular model,  the bonds come in particular lengths, neighboring bonds adopt  particular angles, and these numbers actually matter.    Thus, there were equations, but they were embedded in this structural knowledge.\footnote{One does need equations to compare the predicted structure with X--ray diffraction data,  and these were derived by Cochran, Crick, and Vand \cite{cochran+al}.}  Still, you might worry. 

Second, this was theorizing in which the relevant  principles  were at the level of   molecular structure.  This is a level at which, I think, nobody would doubt that physical principles are relevant for biology. But it isn't clear how you would ever get from that level up to the level that concerns many of us today, the level of ``systems,'' whether we mean systems inside one cell, in a developing embryo, in a network of neurons in the brain, or in a group of organisms behaving cooperatively.  At the opposite extreme, the physical principles to which Rayleigh appealed were completely outside the organism---too macroscopic to help us with most of what we're trying to do today, while what Watson and Crick were doing was too microscopic. Thus, while these examples   tell us that theorizing in this spirit can be incredibly powerful,  the kind of theories that these guys were building doesn't match what we'd like to do today.\footnote{All the talks at the symposium where these ideas were first presented concerned such system level problems (see note 1).  I'll take it as obvious that we will not get from the level of molecular structure up to the level of system organization by detailed simulation, although this itself is a deep issue \cite{anderson_72}.} 

Finally, there is a question about the connection between theory and experiment.  By the time of Rayleigh's work, there was a well established tradition of trying to make quantitative connections between our perceptions and the properties of the physical signals at the input to our sense organs; this subject of ``psychophysics'' would grow and deepen throughout the twentieth century.  The fundamental prediction made by Watson and Crick was about the structure of a molecule, and the decades following their work would see the emergence of X--ray diffraction experiments with atomic resolution, even in large biological structures.  Thus, in both our examples, the theory pointed toward experiments that could be done quantitatively, indeed with methods that are not so far from the traditional core of experimental physics.  Is this the norm, or an exception?

In seminars one often hears words to the effect that  ``the agreement between theory and experiment isn't perfect, but, well, you know, it's biology.''  As an excuse for a little scatter around predictions this might be acceptable, although I  find it a bit annoying.\footnote{It is  a great triumph that, in most areas of physics, such excuses are now unacceptable.  But in many subfields of physics, the expectation of quantitative agreement between theory and experiment is a relatively recent development.  There is much to say about all this, but let me just note that if you don't hope for quantitative agreement, you won't push for it, and as far as I know it never happens by accident.} But in these excuses I sometimes detect an implicit claim that there's more going on, that  there is something fundamental about biology that  prevents us from having the kind of tdetailed, quantitative comparison between theory and experiment that we are used to in the physical sciences.   This is not about describing things at the second decimal place;\footnote{The community of physicists interested in the phenomena of life is not particularly enriched for the kinds of theorists who enjoy making predictions to the second decimal place, or beyond.  But when our  colleagues who {\em are} interested in the second decimal place go after it, they get it right, and this success is part of the reason that we have license to proceed in the ambitious style that we do. So there's a relationship between physics being very precise and being conceptually grand.  Were precision impossible, I suspect that some of the grandeur would be lost. }   the worry is rather that there might be some irreducible sloppiness that we'll never get our arms around, and that this could  spell doom for the physicist's dreams. 

I am surprised by how many physicists simply accept the claim that biology is a messy business.  As explained at (perhaps too much) length in Ref \cite{bialek_12}, one's views on these matters depend on how you are introduced to biology.  If your first exposure is to very complex systems where it is difficult both to maintain control and to make quantitative measurements, then the search for precision can seem hopeless.  But if you start, instead, by studying the ability of the visual system to count single photons, and realize that in the receptor cells of the retina there are $\sim 1\%$ changes in the concentration of internal messengers which are biologically meaningful, you have a different view.   

To summarize, the classical examples are inspiring, but the challenge for theory in our time is (at least) three fold.  First, we have to identify principles that organize our thinking at a systems level.  Second, we have to express these principles in mathematical terms.  Third, if we expect our mathematical theories to make quantitative predictions, we have to push our experimentalist friends to expand the range of life's phenomena that are accessible to correspondingly quantitative measurements.

\section{Hodgkin and Huxley, and beyond}

As a first step in addressing the three problems I have just raised, let me try another classic example, which then moves toward the problems that we want to do today; the classical piece is the work  of Hodgkin and Huxley  \cite{hodgkin+huxley_52}.   They  showed that the electrical dynamics of a neuron---that is, the voltage across the membrane, as a function of space and time---are determined by the dynamics of what we now call ion channel molecules  in the cell membrane.  These are proteins,   and  Hodgkin and Huxley  described the kinetics with which these proteins switch among different states.  This switching depends on the voltage across the membrane, some of the states are open and allow ionic current to flow, others are closed and do not; when you put all of this together,  you end up with a coupled system of nonlinear equations for both the states of the channels and for the voltage itself.  These are the Hodgkin--Huxley equations.

Hodgkin and Huxley studied the squid giant axon, which really is giant, the size of  a small drinking straw. In particular you can pass a wire down the middle of it and short--circuit things so that the voltage all across the membrane is uniform, isolating the dynamics of  ions flowing across the membrane. Once you characterize the dynamics of this ``space clamped'' axon,   you can add back the flow of current along the axon, because this just involves the conductivity of the ionic solution, nothing fancy about the membrane.  The resulting equations predict that  signals converge onto stereotyped pulses that propagate with a definite velocity.  These pulses are the action potentials or ``spikes'' that are the nearly universal mechanism of communication among neurons in the brain \cite{spikes}.  The Hodgkin--Huxley equations predict, correctly, the shape of the voltage spikes and their propagation velocity.

Hodgkin and Huxley had the good fortune that the dynamics of the squid giant axon is dominated by two kinds of ion channels: one sodium channel and one potassium channel. In contrast, our genome encodes roughly one hundred ion channels---more, if you count splicing variants. And the typical neuron in your head might express ten different channels. There was a period which was very productive, during which many groups showed  that the vision of Hodgkin and Huxley was correct, even if most neurons are more complicated than the squid axon:  if you take your favorite neuron, you can reduce its electrical dynamics to a description in terms of several different kinds of channels.  In favorable cases you can even measure, independently,  the current flowing through single ion channels,  watching the channels open and close, and showing that the kinetics of these transitions is consistent with the form of the equations that Hodgkin and Huxley wrote down.    

The industry of building (generalized) Hodgkin--Huxley models for neurons hummed along for decades, resting on a foundation of detailed, quantitative experiments. There is, however, a problem, first emphasized by Larry Abbott and his colleagues \cite{lemasson+al_93,abbott+lemasson_93,goldman+al_01}:  if we are going to use many  different kinds of  ion channels to describe the dynamics of a single neuron, how many of each kind should we use?   While it is possible to measure the kinetics of the individual channels, it's much harder to count directly how many of channels of each type are present in the membrane of a single cell. So, you typically have to extract these numbers by fitting to data on the electrical dynamics itself. Somebody gives you the ingredients and then you have to dial the knobs  to reproduce the behavior of the neuron, and the more realistic the description, with more different kinds of channels, the harder this problem becomes.  The key insight was to step away from the problem of describing a particular neuron and ask:  in the space of all the possible neurons I could build with this many kinds of  ion channels, what can I get? 

\begin{figure}[t]
\begin{centering}
\includegraphics[width = \linewidth]{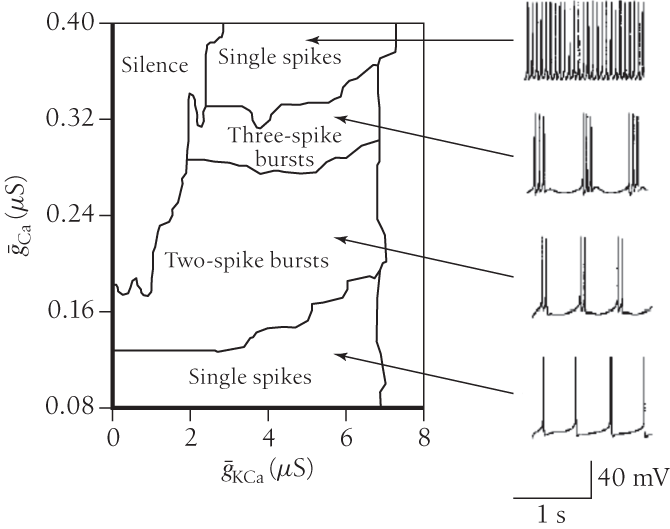}
\end{centering}
\caption{Simulations of a detailed model, with seven types of channel,
for a single neuron in the stomatogastric ganglion
of the crab. Changes in the pattern of activity are shown
as a function of the numbers of two different kinds of
channel, where channel number is expressed as the maximal
conductance when all channels are open.  As explained in the text,  relatively small changes in these parameters can generate   qualitative changes in the pattern
of electrical activity, running the full range from silence to
single spike firing to bursting. After Ref \cite{bialek_12}, redrawn from LeMasson et al \cite{lemasson+al_93}.\label{LFA}}
\end{figure}

An example of the range of possibilities that a cell could access by changing the numbers of just two ion channels is shown in Fig \ref{LFA}.  This example is drawn from a detailed model for a particular neuron in the stomatogastric ganglion of the crab, which has been studied extensively and thus provides a good test case, but there is no reason to think that what we are seeing is specific to this neuron.  By varying the copy numbers of just two types of channel,  we can produce cells that are silent, cells that fire single, isolated action spikes like the ticks of a clock, cells that generate bursts with two or three spikes per burst, and more.  Along one direction we can see transitions through three qualitatively distinct behaviors when the number of copies of one channel is changed by just 10-20\%.  It doesn't take much imagination to think that these quantitative changes in tempo of activity matter to the organism.  This means that our problem in fitting models can be identified with the cell's problem in controlling it's own behavior:  how does a cell manage to sit in the middle of one functional region, and not wander off into other regions?

What Abbott and colleagues proposed was that  cells set the number of channels by  monitoring what they are doing.  So, for example, a cell could monitor it's internal calcium concentration.  When the voltage across the membrane changes, as during an action potential,   calcium channels open and close, calcium flows in, and this  provides  a monitor of  electrical activity. The calcium concentration is known to feed into many biochemical pathways inside the cell, and we can imagine that some of these could regulate either the expression of the channels or their insertion into the membrane.  Mechanisms of this type allow cells to stabilize the very different behaviors seen in Fig \ref{LFA}, essentially because the map of calcium concentration vs channel copy numbers neatly overlays the map of spiking rhythms.  One can do even more subtle things by having multiple calcium sensors with different timescales \cite{liu+al_98}, and much of what we are saying here about the nature of the mapping between channel copy numbers and functional dynamics in single neurons can be generalized to thinking about small networks \cite{prinz+al_04}.

These ideas were quickly confirmed \cite{turrigiano+al_94}.   Perhaps the most dramatic experiment involves taking a neuron,  ripping it out of the network  and putting it in a dish in which the external ionic concentrations are completely bizarre. As a result, when one channel opens the current might flow in the wrong direction, and of course the cell goes completely wild.   But if  you come back a day later, the cell is back doing its normal thing.  It knows what it's trying to do, if one can speak anthropomorphically.  This is a beautiful subject,  still under rapid development.

This example points to a very important transition, which we might think of as   a transition between models and theories. Hodgkin and Huxley proposed a model, and for 30+ years, the goal was to fit that model to the behavior of   particular neurons. It was only in the early 90s that Abbott and company suggested that we look beyond the particular, and take  the generalized Hodgkin--Huxley equations seriously as a theory of what neurons might do.  These equations allow the construction of  cells that belong to a large class, and within that class there are cells that don't exist in nature. Thus, it's not a model of anything in particular, it's a theory for a class of things that can happen, and within that theory there are questions such as how one should set the (many) parameter values.  Stated this way, the question is internal to the theory, but then we can jump to suggest that this is a problem that neurons themselves actually need to solve.   Happily, following this path leads to immediate, and successful, predictions for new experiments.

\section{Problems with parameters}
\label{tuning}

When we make models for the dynamics of a biological system, there are many parameters. In some cases these parameters are encoded in the genome, and change only on evolutionary time scales, while in other cases the parameters are subject to control on physiological time scales, as with ion channel copy numbers.  The more realistic  our models, the more parameters we have, and considerable mathematical ingenuity has been deployed in estimating these parameters from experimental data.  But this whole picture is unsettling for a theoretical physicist.    

Our most complete theories of the natural world certainly have parameters \cite{cahn_96}, but there is a sense that if we are focused too much on these parameters then we are doing something wrong.  If parameters proliferate, we take this as a sign that we are missing some additional level of unification that could relate these many parameters to one another; if our qualitative explanation of phenomena hinges on precise quantitative adjustment of parameters, then we search for the hidden dynamics that could make this apparent fine tuning happen more naturally.  Some of the greatest triumphs of modern theoretical physics are nearly free from parameters---the BCS theory of superconductivity \cite{bcs}, the renormalization group theory of critical phenomena \cite{wilson_75},  the theory of the fractional quantum Hall effect \cite{laughlin_83}, and more.  Importantly, these examples refer not a rarefied world of interactions among small numbers of elementary particles, but rather to the properties of real, macroscopic materials, with all their chemical complexities.\footnote{There are examples in the same spirit from elementary particle physics, notably the connection of deep inelastic scattering experiments to the asymptotic freedom of QCD \cite{gross+wilczek_74}.}  

How can we reconcile the parameter aversion of theoretical physicists with the explosion of parameters that arise in a realistic approach to biological systems?  Much of what our community is doing, I think, can be understood as a reaction to this problem.  There are several approaches.\footnote{One possibility, surely, is that the multitude of parameters is a fact of life, and somehow irreducible, in which case we need to give up on our search for a physicist's understanding. I'll discard this as too pessimistic.}  First, it might be that the parameters are just a distraction, and that the  meaningful functional behaviors of biological systems emerge as generic or ``robust'' properties of our models, independent of precise parameter settings.  A second, approximately opposite view is that the forces of evolution have been strong enough to select non--generic parameter values, allowing for phenomena that emerge only through fine tuning; if we can identify the selection principle, we then have a theory for at least an idealization of the real biological systems, again without reference to parameters.  Finally, we might hope that parameter independence emerges in biological systems much as it does for inanimate materials, with something like the renormalization group telling us that macroscopic behaviors which matter for the organism can be independent of (highly parameterized) microscopic details.    In these three sections (\S\S\ref{tuning}--\ref{collective}), I'll look at these three ideas in turn.

The problem discussed in the previous section is exactly the problem of balancing robustness against fine tuning; interestingly, the picture proposed by Abbott and colleagues  essentially splits the difference between these very different ideas.  Within a description of ion channel dynamics alone, what we see in real neurons is manifestly the result of fine tuning:  you have to get combinations of  ion channel copy numbers right, as shown in Fig \ref{LFA}.  But the mechanism by which cells achieve this tuning is to promote the finely tuned parameters to being dynamical variables,\footnote{Promoting  a finely tuned parameter to a dynamical variable is natural in the biological context, where  there are many layers of regulation.  It seems worth noting that  the same strategy was suggested in the context of particle physics for solving the `strong CP problem,' and in that case the new dynamical variables are associated with a new (still hypothetical) elementary particle, the axion \cite{axion}.} and then this larger dynamical system can be attracted to a functional fixed point, generically.  It also is important that the mapping from parameters to function is complex, and many--to--one.

In the conventional language of neuroscience, the parameters generalized Hodgkin--Huxley models are measured as ``maximal conductances'' for each type of channel, that is the conductance  if all the channels of  a particular type are in their open state.  This is the product of a single channel conductance and the number of channels, so I have referred to this as a problem of ion channel copy numbers;  since ion channels are proteins, this problem is about how the functional dynamics of a network of interacting proteins depends on the number of copies of each protein.   This problem re--emerged some years later in thinking about biochemical and genetic networks \cite{barkai+leibler_97,dassow+al_00}, where it grew into a separate literature.\footnote{It really does need to be emphasized that the Hodgkin--Huxley model and its descendants are models for  networks of interacting proteins (the channels), where the interactions are mediated through membrane voltage.  Perhaps the most important  contrast to most other such networks is that, for ion channels, we actually know the equations that describe the relevant dynamics.  In the non--neuronal examples, one can write schematic equations based, for example, on Michaelis--Menten enzyme kinetics, but often many qualitative features of the dynamics are unknown.  These differences in mathematical description are tied to differences in experiments, since in neurons it has long been possible to make high precision, real time measurements of the electrical signals that form the functional output of the ion channel network in single cells, while similarly precise measurements on biochemical and genetic networks remain challenging.}

In biochemical and genetic networks, there has been considerable emphasis on the need for 
``robustness'' against protein copy number variations.   This idea resonated in the community in part because of a shared, if implicit,  hypothesis that precise control over protein copy numbers is not possible.  While some systems might indeed be robust, I think we now now that precise control of protein copy number is, in fact, possible when needed.  The ion channel example shows that copy number fluctuations can be large, but the functionally important combinations of copy numbers can be tuned through feedback.   In contrast,  the example of maternal morphogens in the fruit fly embryo shows that cells can generate reproducible copy numbers even without feedback, so that absolute concentrations can carry biologically meaningful signals \cite{gregor+al_07,petkova+al_14}.    In  bacterial chemotaxis, which provided some of the motivation for the robustness idea \cite{barkai+leibler_97}, more recent experiments show that the operon structure of gene regulation in bacteria serves to reduce relative fluctuations in the copy numbers of crucial proteins, and that if this structure is removed then cells that exhibit larger relative fluctuations are at a competitive disadvantage \cite{kollmann+al_05,lovdok+al_09}.  It thus seems likely that, in all these systems, we are seeing tuning or selection of parameters to achieve functional outputs, and that the search for networks which can achieve functionality with random parameter choices may be missing something essential.

From the level of interacting networks of proteins we can drop down to ask about genericity vs. fine tuning in single protein molecules.  We recall that proteins are polymers of amino acids, with lengths from a few tens to many hundreds of residues, and with twenty types of amino acids the number of possible proteins is (beyond) astronomical.  The sequence of amino acids in most cases determines the structure, and hence the function, of the protein.  On the one hand, we know that these molecules are not  finely tuned:  not every single detail of the amino acid sequence matters. On the other hand,  a random sequence typically doesn't even fold into a unique compact structure---random heteropolymers are glassy---let alone carry out interesting functions.  So where along the continuum between every detail being important to being completely generic do real proteins sit?   We will return to this problem below.

Instead of moving down to the level of single protein molecules, we can look for examples of this same question by moving ``up'' to the level of neural networks. In particular, let's think about the problem of building a short--term memory for a continuous variable. If I want a network that generates the pattern of activity needed to hold my arm fixed at each of several different heights, then I need a dynamical system that has a fixed point at each of these locally stable positions.  But if I want to hold stable at a continuous range of positions, I need a line of fixed points, and that's completely non--generic. One fixed point is okay,  and multiple isolated fixed points are fine, as in the Hopfield model \cite{hopfield_82}, but a whole continuum of fixed points---that's not generic, you have to tune parameters.

Better than the problem of holding your arm at a fixed height (which involves feedback from mechanical sensors) is the problem of   holding your eyes still.   With your eyes open you have visual feedback, but  even if you close your eyes and turn your head you still counter--rotate your eyes to compensate for the movement.  The signal coming from your ears (more precisely, from the semicircular canals) is a motion signal, but in order to keep your eyes counter--rotated you need a position signal. So you integrate and hold onto the result after the inputs have disappeared.   We can do this for times on the order of a minute, whereas individual neurons usually forget their inputs over perhaps tens of milliseconds. So you have a gap to span, across several orders of magnitude in time.  

Notice that there are two theoretical ideas here \cite{seung_96}.  First,  we should think about something as (seemingly) simple as holding our eyes still in terms of networks with a line of fixed points.  Second,  in the space of possible networks of neurons, such behavior is  not generic, so one needs an explanation of how it can occur.

A natural answer to the problem of stabilizing non--generic behavior is that since this is a brain, it can learn.  In fact, the brain has constant access to a feedback signal: if you fail to get things right, then the world keeps slipping on your retina.  So the brain should be able to exploit this signal and tune the relevant network, somehow, to achieve this very non--generic dynamics. If this picture of tuning via feedback is correct, and we disrupt the feedback, we should be able to ``un--tune'' the system.  There's a beautiful experiment by  David Tank and colleagues showing that this is true \cite{major+al_04a,major+al_04b}. 

\begin{figure}[b]
\begin{centering}
\includegraphics[width = \linewidth]{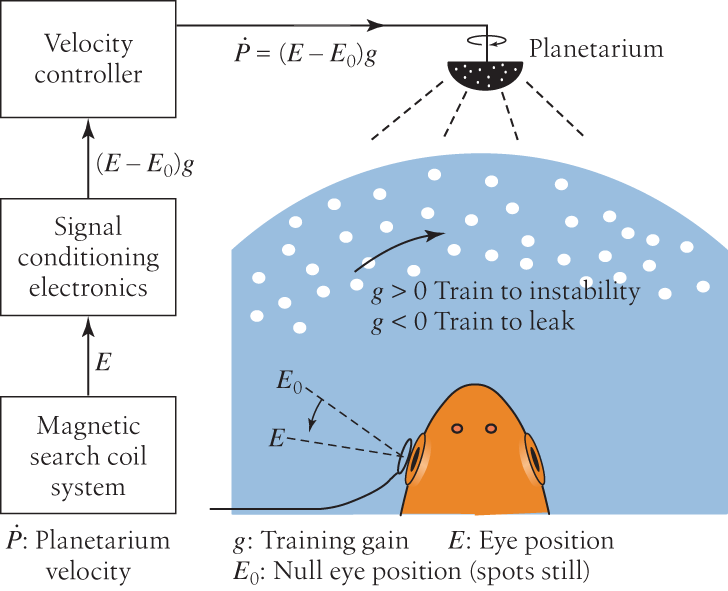}
\end{centering}
\caption{Schematic of the ``planetarium'' experiment. Eye movements are monitored and fed back to movements of the surrounding scene with an adjustable gain. After Ref \cite{bialek_12},  redrawn from Major et al \cite{major+al_04a}. \label{DWT1}}
\end{figure}

\begin{figure}[t]
\begin{centering}
\includegraphics[width = \linewidth]{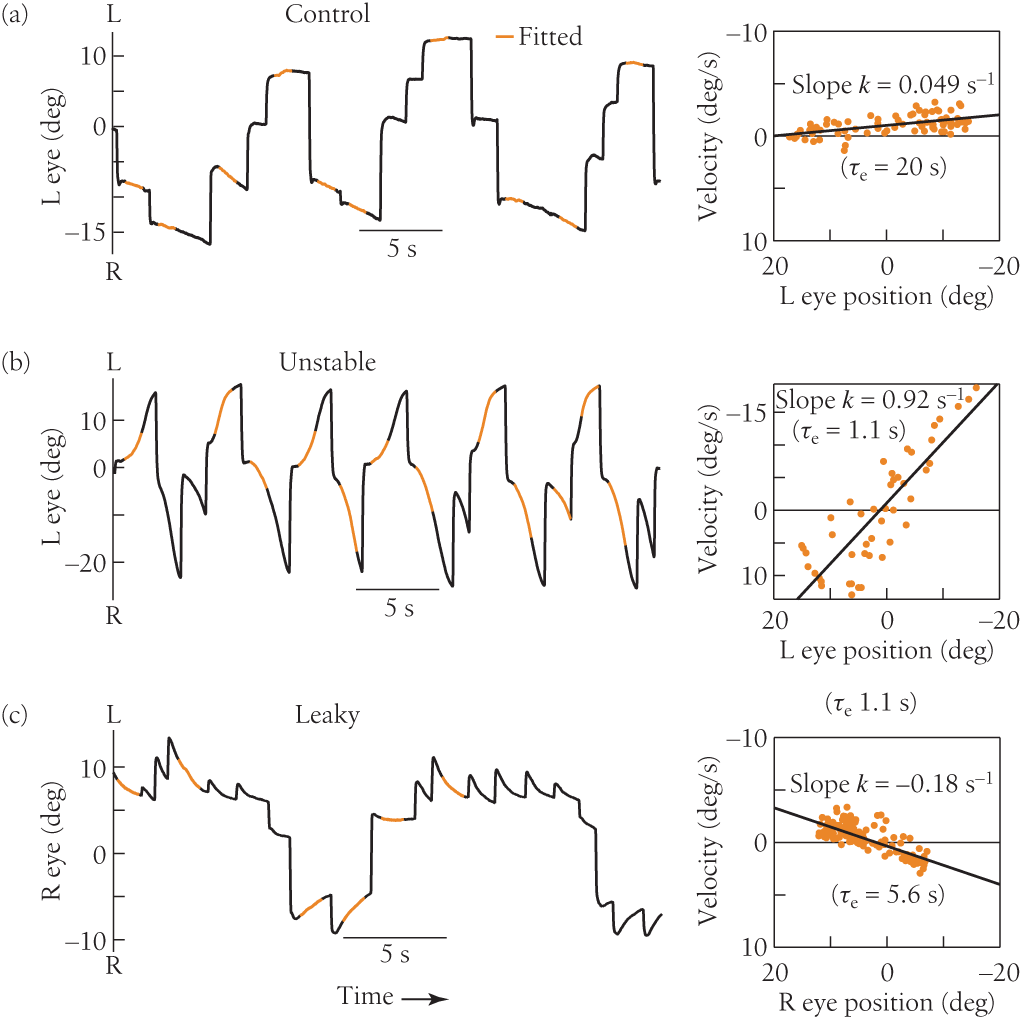}
\end{centering}
\caption{Results of the planetarium experiment.  At left, eye position
versus time. At right,  velocity versus position along these trajectories. (a) Control experiments
 before exposure to the feedback system in the planetarium. Note that the time constant of the system is $\sim 20\, {\rm s}$. After exposure to feedback, which should ``teach'' the system to be unstable (b) or leaky (c), measurements show the expected behaviors, with time constants for growth or decay on the order of $1- 5\,{\rm s}$.  After Ref \cite{bialek_12},  redrawn from Major et al \cite{major+al_04a}.\label{DWT2}}
\end{figure}

The essence of the experiment, as schematized in Fig \ref{DWT1}, is to build a planetarium for goldfish, a seemingly low tech experiment that gets right at the central question.  This setup monitors the motion of the eyes of the goldfish and rotates the world in proportion,  thus  changing the coupling between the eyes' rotation and the world's rotation.  And so if the brain is tuning the networks  that stabilize eye movements using visual motion signals, placing the fish in this apparatus  will cause the system to mistune.   After learning in this unusual environment, the network won't  hold a constant eye position, but will be either  unstable, with the eye being driven off to eccentric positions, or   leaky, with eye positions  relaxing back to the middle.  This is exactly what is seen experimentally (Fig \ref{DWT2}, \cite{major+al_04a}), and one can even trace these changes in stability down to the dynamics of individual neurons in the network \cite{major+al_04b}.  This  shows  that you actually have active tuning mechanisms which  stabilize this very non--generic behavior of the underlying dynamical system.

\section{Signals, noise, and information}
\label{info}

The mechanisms of life can achieve extraordinary precision \cite{bialek_12}:   our visual system can count single photons, we can hear sounds that cause our eardrum to vibrate by the diameter of an atom, bacteria swim along the gradients of attractive chemicals with a reliability so high that they must be counting every molecule that arrives at their surfaces.   I have been interested for some time in whether these are isolated examples, or whether biological systems more generally operate near the limits of what is allowed by the laws of physics.\footnote{Given the task of providing some perspective on the problem of theorizing about biological systems, I cannot resist discussing problems that my colleagues and I have been thinking about.  There are two themes, which we have tried to follow across many different systems, as described in this Section and the next.  Evidently the opinions I state here have much less claim to objectivity than in my description of other people's work.}  If (near) optimality is the rule, then we can promote this to a principle from which essential aspects of the underlying mechanisms can be derived, quantitatively.  In some cases the resulting theoretical structure is naturally phrased in terms of optimizing the reliability of decisions, or the accuracy of estimates, while in other cases it seems more compelling to use the slightly more abstract framework of information theory.  In either case, I think it's crucial that we not adopt sweeping hypotheses of optimality for aesthetic reasons, but try to focus on examples where the approach to optimality can be tested, directly, through quantitative experiments.

Like us, flies use their visual systems to help guide their movements.  But, flying at meters per second, they are under pressure to make very quick decisions, and looking out through the tiny lenses of the compound eye, the raw data they have to work with has rather low resolution; this combination of physical constraints means that even optimal visual estimates of how they are moving through the environment may not be so reliable.  Estimates of motion, in particular, are encoded by sequences of action potentials from a relatively small number of neurons deep in the fly's brain; wide--field, or rigid body motions are the responsibility of some rather large neurons, and even thirty years ago it was possible to make very long, stable recordings from these cells \cite{ruyter+al_86}.   At the same time, one can calibrate the signal and noise properties of the photoreceptors, showing that these are nearly ideal photon counters, albeit with finite time resolution, up to counting rates of $\sim 10^5/{\rm s}$ \cite{ruyter+laughlin_96a,ruyter+laughlin_96b}.  Rob de Ruyter van Steveninck and I worked together to show that the motion--sensitive neurons encode estimates of visual motion with a precision within a  factor of two of the limits set by receptor cell noise and diffraction blur \cite{bialek+al_91,ruyter+bialek_95}.   

The observation that the fly can make motion estimates with a precision close to the physical limits suggests that a theory of optimal estimation might be a theory of the computations actually done by the fly's brain.  We have developed this theory \cite{potters+bialek_94}, and found signatures of the predicted behavior in the responses of the motion--sensitive neurons \cite{ruyter+al_94,bialek+ruyter_05}, but it must be admitted that the jury is still out.  We have used similar arguments to derive the filtering characteristics of the first synapse in the retina, optimizing the detectability of single--photon signals \cite{bialek+owen_90,rieke+al_91};  this may have been the first example of using optimization arguments to generate successful parameter--free predictions of neural responses.  Subsequent work has explored the  role of nonlinearities in separating single--photon signals from noise at this synapse \cite{field+rieke_02}, and there have been efforts to use optimization arguments to understand aspects of visual motion perception in humans \cite{weiss+al_02,stocker+simoncelli_06}.\footnote{Human perception is a rich source of quantitative data, as noted above in connection with Rayleigh's classic work.  But, in contrast to the example of fly vision, it can be difficult to calibrate the noise levels at the input to the human visual system, except in the limit where we are counting single photons.  Thus, while it has become popular to use optimality arguments in relation to human perception, I think there have been fewer direct tests of optimality than one might like.}  The case of visual motion estimation in flies is receiving renewed attention \cite{fitzgerald+al_11,clark+al_14,fitzgerald+clark_15}, in part because of opportunities to combine genetic and structural tools to dissect the layers of circuitry that lead from the receptor cells to the larger motion sensitive neurons \cite{takemura+al_13,fisher+al_15}.

Organisms must respond to changing concentrations of molecules in their environment, and many internal signals are encoded by such concentration changes.  As first emphasized by Berg and Purcell in the context of bacterial chemotaxis, there is a physical limit to the precision of such signaling because the relevant molecules arrive randomly at their targets, creating a form of shot noise  \cite{berg+purcell_77}.  My colleagues and I have tried to make the intuitive arguments of Berg and Purcell more rigorous, with the goal of defining limits to  signaling in a broader range of biological processes \cite{bialek+setayeshgar_05,bialek+setayeshgar_08,tkacik+bialek_09}, and this problem has now been addressed in several different ways \cite{endres+wingreen_09,mora+wingreen_10,govern+wolde_12,kaizu+al_14,paijmans+al_14}.    We have worked with our experimental colleagues to show that the limits are reached, or at least approached, in the early events of embryonic development in the fruit fly, as the network of gap genes responds to spatially varying concentrations of the primary maternal morphogens \cite{gregor+al_07,tkacik+al_08}.

A more abstract notion of optimal performance concerns the efficiency of information transmission and representation, an idea that reaches back to discussions of neural coding, perception, and learning by Barlow and Attneave in the 1950s \cite{barlow_59,barlow_61,attneave_54}.  The ability of neurons to convey information is limited by the statistical properties of the action potential sequences that they generate, and by the time resolution with which the brain can meaningfully `read' these sequences \cite{mackay+mcculloch_52}.    We have worked with experimental collaborators to show that real neurons transmit information about dynamic sensory inputs at rates within a factor of two of the physical limit set by the entropy of the spike sequences, down to time resolutions on the order of milliseconds \cite{rieke+al_93,strong+al_98a,strong+al_98b}, and that this efficiency is even higher for inputs that capture some of the statistical features of the relevant natural signals \cite{rieke+al_95,lewen+al_01,wright+al_02,nemenman+al_08}.  As first emphasized by Laughlin, such efficiency requires a matching of neural coding strategies to the statistical structure of sensory inputs \cite{laughlin_81}.

\begin{figure*}[t]
\begin{centering}
\includegraphics[width = 0.9\linewidth]{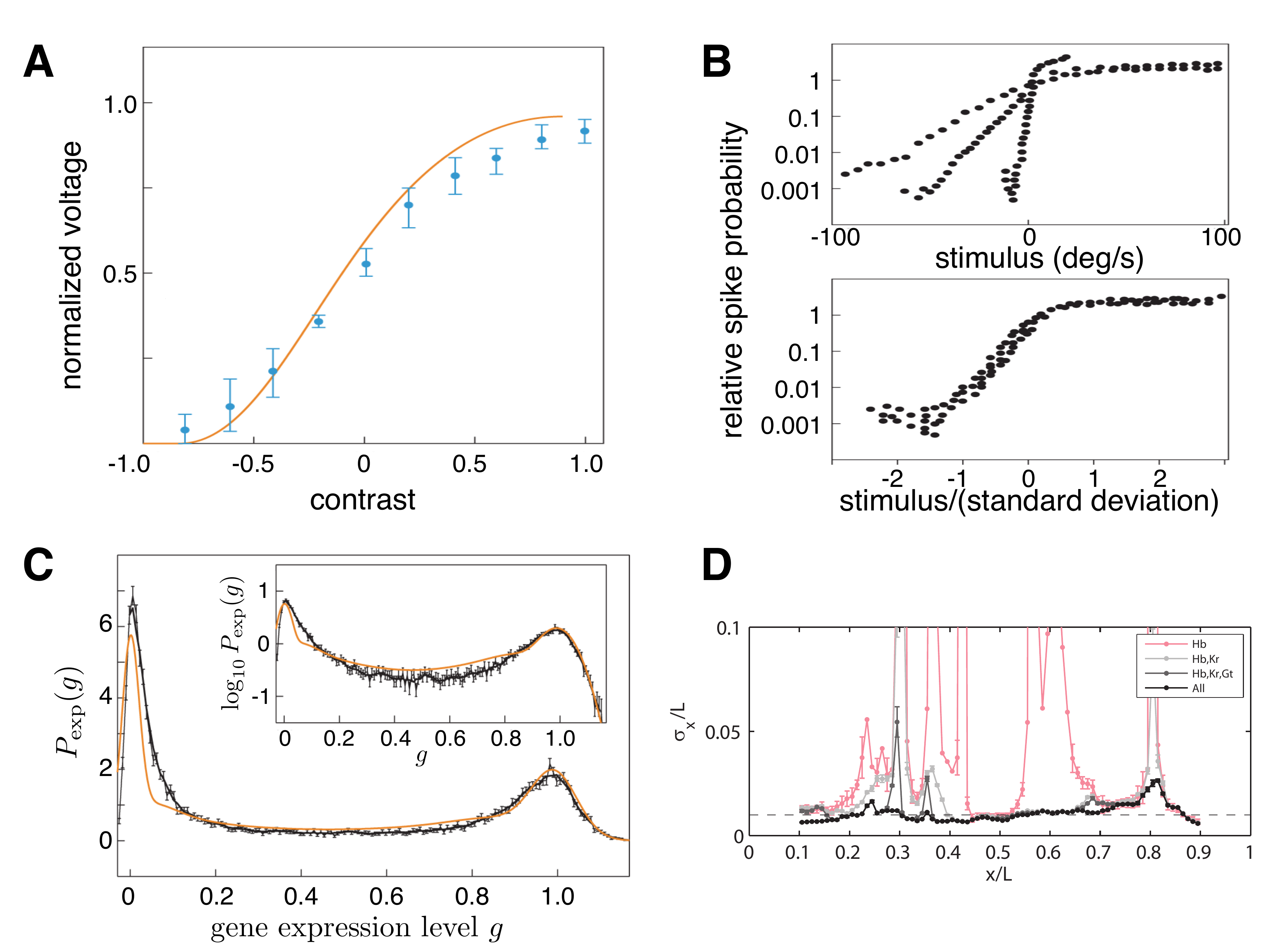}
\end{centering}
\caption{Matching the distribution of inputs to the characteristics of a communication channel. (A) Voltage responses of the large monopolar cells (LMCs) in the fly retina to step changes in light intensity (points with error bars), compared with  the cumulative distribution of image contrast in the fly's environment (solid line) \cite{laughlin_81}.   (B) Spiking responses of the motion sensitive neuron H1 in the fly visual system \cite{brenner+al_00a}.  Input/output relations are different when measured with input distributions that have different width (top), and  collapse under rescaling by the dynamic range of the inputs (bottom).   (C) Expression levels of Hunchback in the early {\em Drosophila} embryo \cite{tkacik+al_08b}.  Solid line is the optimal distribution of outputs calculated from measured noise levels in the transformation from Bicoid inputs to Hunchback outputs, and points with error bars are the data; inset shows the distributions on a logarithmic scale.  (D) Precision of positional estimates ($\sigma_x$) based on  the expression of all four gap genes (Hb, Kr, Gt, Kni) in the  {\em Drosophila} embryo \cite{dubuis+al_13}.   \label{matching}}
\end{figure*}

Laughlin considered the response of large monopolar cells (LMCs) in the fly retina to changing image intensity or contrast.  These cells give a graded voltage output, and in the limit that voltage noise is small and independent of the mean, the optimal input/output relation is one that generates a uniform distribution of outputs; this means that the normalized input/output relation is the cumulative distribution of inputs.  Laughlin built a photodetector with optics matched to that of the fly's eye and sampled this distribution in the natural environment.  He then compared the predictions with the voltage responses of the LMCs, as  in Fig \ref{matching}A.   These results inspired explorations of this ``matching''  principle  in different contexts.

Since natural signals are intermittent, optimizing information transmission requires that sensory neurons  adjust their input/output relations in real time as the dynamic range of inputs varies \cite{ruderman+bialek_94}.   These effects were demonstrated in the vertebrate retina \cite{smirnakis+al_97} and in the motion sensitive neurons of the fly visual system \cite{brenner+al_00a}.  In particular, if we consider a family of input distributions that differ only in the overall dynamic range of inputs (e.g., Gaussian signals with different variances), then when the noise  is small the only parameter that can set the scale of inputs is the dynamic range itself.  Hence optimal input/output relations in these different environments should be rescaled versions of one another, as seen in Fig \ref{matching}B \cite{brenner+al_00a}; one can even show that the proportionality constant between the gain of the input/output relation and the dynamic range of input is the one that maximizes information transmission.  Related adaptation effects have now been seen in a wide range of systems, at levels from the sensory periphery to deep in the cortex \cite{kvale+schreiner_04,dean+al_05,nagel+doupe_06,maravall+al_07,debaene+al_07,wen+al_09,dahmen+al_10,rabinowitz+al_11}, and there are even hints that the speed of adaptation itself approaches the limits set by the need to collect reliable statistics on the input distribution \cite{fairhall+al_01,wark+al_09}.  

Limits to information transmission are set by a combination of the available dynamic range and the noise levels in the signaling pathway.  These limits are especially clear when the signals are carried by changes in the concentration of signaling molecules; an important example is transcriptional regulation, which we can think of as the transmission of information from the concentration of transcription factors to the expression levels of the target gene(s) \cite{tkacik+al_08a}.   In the limit that noise levels are small, but state--dependent, optimizing information transmission leads to a distribution of outputs inversely proportional to the standard deviation of the noise; more generally, if we have a  characterization of the noise level along the input/output relation, we can find the optimal distribution of outputs numerically.  In the {\em Drosophila} embryo, the expression level of Hunchback responds to the spatially varying concentration of the primary maternal morphogen Bicoid, and from the measured noise levels \cite{gregor+al_07} we can compute the optimal distribution of expression levels, which is in surprisingly good agreement with experiment  \cite{tkacik+al_08b} (Fig \ref{matching}C).   Hunchback is one of several gap genes, and together the expression levels of these genes are thought to provide information about the position of cells ($0 < x< L$) along the anterior--posterior axis of the embryo.     Because the distribution of positions is uniform, matching requires that the errors in estimating position ($\sigma_x$) also be uniform.  If we look at just one gene, this is far from the case, but as we add in the contributions from all the gap genes, with their complicated spatial patterns of mean expression and (co)variance, we see the emergence of a  nearly uniform positional error, as shown in Fig \ref{matching}D \cite{dubuis+al_13}.  Importantly, the scale of this positional error  ($\sigma_x \sim 0.01 L$) is essentially equal to the precision of subsequent decisions about the body plan.

Before leaving Fig \ref{matching}, let me emphasize that in each case we are using the same theoretical principle:  maximize information transmission by matching the distribution of inputs to the input/output relation and noise levels.  There are differences of detail, but these arise because the noise levels in the different systems are different.  Crucially, this theoretical approach generates parameter--free predictions; thus, none of the results in Fig \ref{matching} involve fitting.  Further, in each case we can not only test predictions based on optimizing information transmission, we can also estimate the amount of information being transmitted and show that it is very close to the optimum.\footnote{For cases (B--D) in Fig \ref{matching}, this point is made in the cited papers.  For the case of the LMCs, see Ref \cite{ruyter+laughlin_96b} and \S3.1 of Ref \cite{spikes}.}

Beyond matching, we have been searching for the architecture and parameters of genetic networks that optimize information transmission, and we have been able to formulate this optimization problem in a way such that the solutions depend only on the number of available molecules.  As a function of this resource constraint, we find transitions from architectures that are highly redundant, with multiple target genes responding identically to transcription factor inputs, to architectures where the multiple targets are activated or repressed at staggered thresholds, tiling the dynamic range of inputs \cite{tkacik+al_09a}.  Redundancy can be reduced, and efficiency increased, by mutual repression among target genes \cite{walczak+al_10}, feedback loops can generate long integration times to help average out noise \cite{tkacik+al_12}, and in spatially extended systems such as a developing embryo the proper amount of spatial averaging can play a similar noise--reducing role \cite{sokolowski+al_15a}; finally,  the cell can enhance information transmission  at the lowest transcription factor concentrations by having these molecules act also as translational regulators of constitutively expressed mRNAs \cite{sokolowski+al_15b}.  All of these theoretical results have qualitative correlates in the properties of real genetic control networks, notably the gap gene network in the developing fly embryo, although it remains  a challenge to put these different results together into a fully quantitative theory of real networks.

In all the examples above, successful application of information theoretic ideas depends on identifying what information is relevant to the  system we are studying.  One can imagine a nightmare scenario in which the very principled notion of optimizing information transmission is submerged under long arguments about natural history, and our hopes for theory in the physics sense are dashed.  Can we do something more general?  
We have argued that, in many cases, information is relevant to the extent that it has predictive power \cite{bialek+al_01,bialek+al_07};  predictive information captures our intuition about the complexity or richness of  time series \cite{bialek+al_01}, and the efficient representation of predictive information unifies the description of signal processing and learning \cite{bialek+al_07}.    In collaboration with MJ Berry II and his colleagues, we have now measured the predictive information carried by neurons in the vertebrate retina \cite{palmer+al_15}.  Every ganglion cell participates in a small group for which the encoded predictive information is close to the  limit set by the statistical structure of the inputs themselves. Groups of cells  carry information about the future state of their own activity, and  this information can be extracted by downstream   neurons that exhibit familiar forms of visual feature selectivity.  The efficient representation of predictive information is a new candidate principle that can be applied at every stage of neural computation.

\section{Collective behavior}
\label{collective}

From the spectacular aerial displays of flocking birds down to the beautiful choreography of cell movements in a developing embryo, many of life's most striking phenomena emerge from interactions among hundreds if not thousands or even millions of components.  The enormous success of statistical physics in describing emergent phenomena in equilibrium systems has led many people to hope that it could provide a useful language for describing emergence in biological systems as well.   In the past decade or so, my colleagues and I have been excited by the use of maximum entropy methods to build statistical physics models for  variety of biological systems that are grounded in real data.\footnote{The Boltzmann distribution is the maximum entropy distribution consistent with knowing the mean energy, and this sometimes leads to confusion about maximum entropy methods as being equivalent to some sort of equilibrium assumption (which would be obviously wrong).  But we can build maximum entropy models that hold many different expectation values fixed, and it is {\em only} when we fix the expectation value of the Hamiltonian that we are describing thermal equilibrium.   What is useful is that maximum entropy models are equivalent to the Boltzmann distribution for some hypothetical system, and often this is a source of both intuition and calculational tools.} 

In a small window of time, a single neuron either generates an action potential or remains silent, and thus the states of a network of neurons are described naturally by binary vectors.  We have tried to approximate the probability distribution of these binary vectors by maximum entropy distributions that are consistent with the mean spike probability for each cell, and with the matrix of pairwise correlations among cells.  These models are Ising models, and since correlations have both signs, the interactions among ``spins'' in the model have both signs---they are a sort of spin glass, not unlike the model that Hopfield wrote down in 1982 \cite{hopfield_82}.  With MJ Berry II and his colleagues, who have developed methods for recording simultaneously from almost all the  neurons in a small patch of the vertebrate retina as it responds to naturalistic visual inputs \cite{segev+al_04,marre+al_12},  we found that  models based on pairwise correlations provided strikingly precise descriptions of the entire distribution of neural activity in groups of ten to fifteen cells \cite{schneidman+al_06}.   By now we can write very accurate probability distributions for the joint activity of 160 cells in the vertebrate retina \cite{tkacik+al_14a}.  Although there are many details, the overall structure of these models is consistent with extrapolations from the analysis of smaller groups of cells \cite{tkacik+al_06,tkacik+al_09}, and aspects of this structure can be seen in much simpler models \cite{tkacik+al_12a}.  We have preliminary evidence that the same maximum entropy strategy can describe activity in populations of $\sim 100$ neurons in the hippocampus \cite{meshulam+al_15}.

Around the time we were getting our first results on maximum entropy models for neurons, I heard Rama Ranganathan talk about his group's efforts to explore the space of amino acid sequences.  In outline, they looked at a family of proteins that were known to have similar structures and functions, and developed an algorithm to generate a new ensemble of sequences that were consistent with the observed pairwise correlations among amino acid substitutions at different sites along the chain.   They then synthesized some of the molecules in this artificial family, and found that a substantial fraction of these molecules were functional; in contrast, proteins synthesized by choosing amino acids independently at each site were not functional \cite{socolich_al_05,russ+al_05}.  We were able to show that what Ranganathan and colleagues were doing was, in a certain limit, equivalent to the pairwise maximum entropy construction that we were doing for neurons \cite{bialek+ranganathan_07}.  

In the maximum entropy construction, correlations between substitutions at different sites are generated by effective interactions, and from other statistical mechanics problems we expect that the spatial range of correlations will be larger than the spatial extent of interactions \cite{lapedes+al_98,giraud+al_99}.  Indeed, one can find correlations among amino acid substitutions that are widely separated, not only along the polymer chain but also in three dimensional space, but our intuition is that interactions should be local.  If this  is borne out, then the statistics of pairwise correlations among amino acids substitutions encodes information about which sites along the one--dimensional sequence are neighbors in three--dimensional space, and  we would be able to predict protein structures from sequence data alone \cite{lapedes+al_02}.   There is tantalizing evidence from Weigt, Colwell, and others that this actually works \cite{weigt+al_09,marks+al_11,sulkowska+al_12}.  These models also provide an explicit answer to the question raised in \S \ref{tuning} about the location of amino acid sequences along the continuum from fine tuning to randomness.

\begin{figure*}[t]
\begin{centering}
\includegraphics[width = 0.9\linewidth]{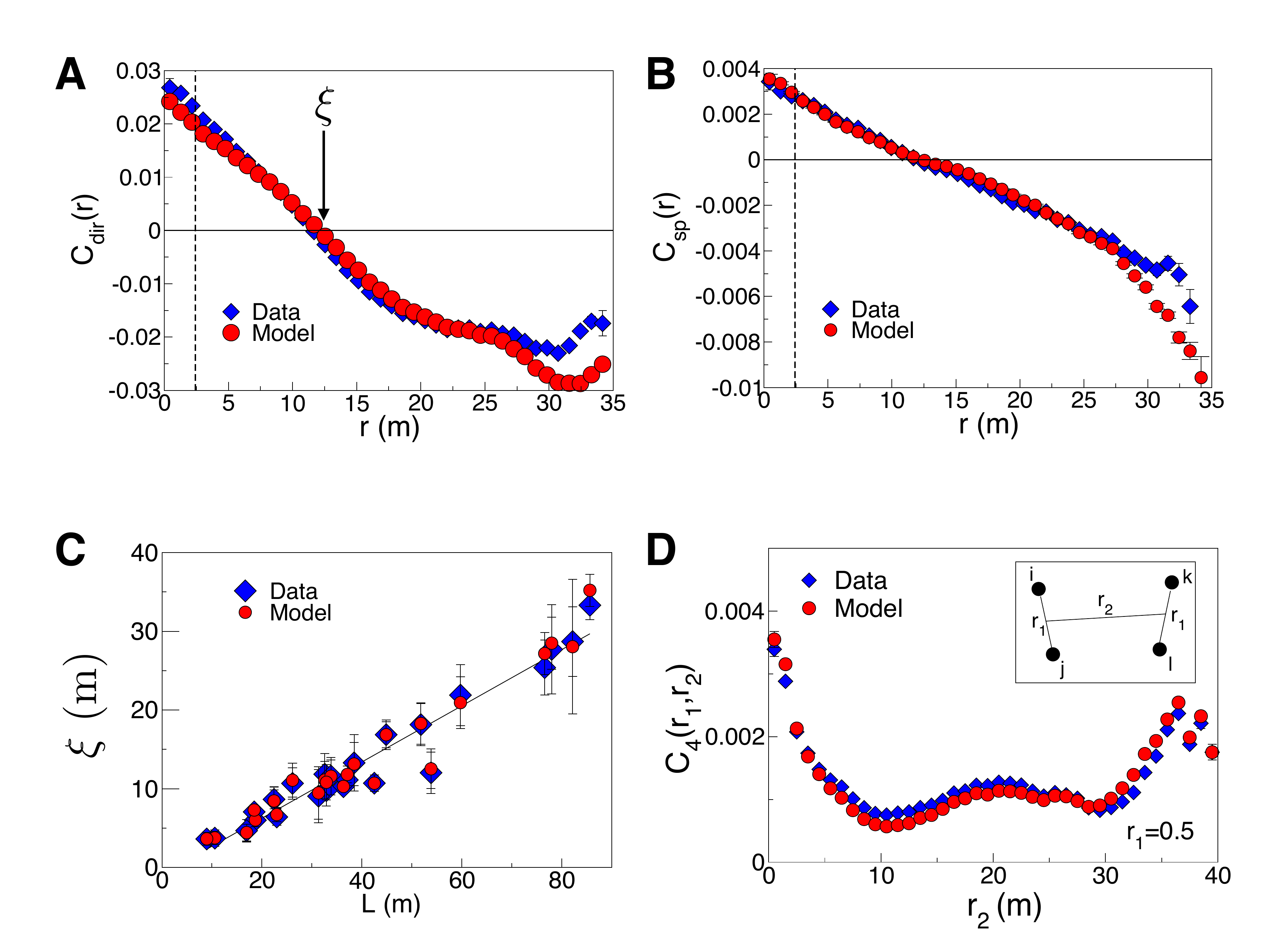}
\end{centering}
\caption{Statistical mechanics for a natural flock of birds, from Refs \cite{bialek+al_12,bialek+al_14}. For each bird, we decompose the vector velocity into flight speed and a unit vector that indicates flight direction,   $\vec{\mathbf v} = \nu\hat{\mathbf v}$.  (A) Connected correlation function of the direction fluctuations, $C_{\rm dir}(r) = \langle (\hat{\mathbf v}_{\rm i} - \bar{\mathbf v}){\mathbf\cdot} (\hat{\mathbf v}_{\rm j} - \bar{\mathbf v})\rangle$, where the average is over all pairs of birds with distance $r_{\rm ij} = r$, and the mean is taken over a single snapshot of the flock, $ \bar{\mathbf v} = (1/N)\sum_{\rm i} \hat{\mathbf v}_{\rm i}$. (B) Connected correlation function of the speed fluctuations, defined similarly, in units where the mean speed is one.  (C) Dependence of the apparent correlation length (from A) on the linear dimensions of the flock.  (D)  Connected four point correlations of the directional fluctuations.  In all panels, red points are from the model and blue points from the data, as indicated.  The model matches the mean and variance of the speed, as well as the average correlation of a bird's velocity with that of its near neighbors; the dashed lines in (A) and (B) indicate the size of this neighborhood; everything beyond this distance is a parameter--free prediction.  \label{birds}}
\end{figure*}

Perhaps the prototypical example of emergent, collective behavior in a biological system is a flock of birds.  There were important early theoretical efforts to develop a statistical mechanics of flocking and swarming \cite{toner+tu_95,vicsek+al_95,toner+tu_98}, and these ideas developed into a whole field of ``active matter'' \cite{ramaswamy_10}, but  I think it is fair to say that, well past the year 2000, most of the experimental observations were qualitative.  The situation changed dramatically with the work of Cavagna, Giardina, and their colleagues in Rome, who developed methods to track the trajectories of every bird in groups of more than one thousand starlings as they engaged in aerial displays  \cite{ballerini+al_08a,cavagna+al_08a,cavagna+al_08b,ballerini+al_08b}.  We have worked together to   build maximum entropy models for the joint distribution of velocities for all the birds in the flock, matching the average correlation of birds with their near neighbors, as well as mean and variance of the speeds \cite{bialek+al_12,bialek+al_14}.  Again, these extremely simple models are strikingly accurate, as shown in Fig \ref{birds}, correctly predicting the pattern of correlations throughout the entire flock, including the small but significant four--bird correlations, as well as the long--ranged correlations  in the fluctuations of both flight direction and flight speed \cite{cavagna+al_10}.      Again,   Fig \ref{birds} is not a collection of fits; the model is determined by matching three local expectation values, one of which simply sets the units of speed, and everything else that we calculate is a parameter--free prediction.  In particular, we are not free to make adjustments in an attempt to capture the long--ranged correlations; either these are predicted correctly, or they are not.

These models are mathematically equivalent to equilibrium statistical mechanics models with local interactions,  and in such systems long--ranged correlations can arise only by two mechanisms:  Goldstone's theorem, and tuning to a critical point. Indeed, the flock spontaneously breaks a continuous symmetry by choosing an overall flight direction, and the long--ranged correlations in the directional fluctuations are mediated by the  resulting Goldstone modes.  But there is no corresponding argument for the speed fluctuations, and in this case long--ranged correlations must be a signature of criticality, as one can verify by detailed analysis of the model in Ref \cite{bialek+al_14}.\footnote{Because the flock is an active system, even if the real interactions among birds are local, the joint distribution of velocities at one moment in time might not be well approximated by a local model.  Thus it is important that, within the maximum entropy framework, one can test for the (un)importance of longer ranged interactions \cite{cavagna+al_15}.  One can go further, and build maximum entropy models that describe the dynamics of the flock, and for real flocks of starlings much of our equilibrium intuition is valid because the time scales for rearrangement of the birds in the flock is much longer than the time scale for equilibration of birds with their neighbors \cite{cavagna+al_14,mora+al_15b}.}     The Rome group has gone on to analyze the trajectories of swarming midges, and here too they see long--ranged correlations of velocity fluctuations, now in the absence of symmetry breaking, and argue that this again is  a sign of criticality \cite{attanasi+al_14a,attanasi+al_14b}.

For neurons, the notion of locality of interactions is not so useful, because neurons are extended objects and can reach many, many neighbors.   As a result, long--ranged correlations are not a useful diagnostic of criticality.   As an alternative we have tried to develop a thermodynamics for neural networks, essentially counting the number of states (combinations of spiking and silence across the population) that have a particular value of log probability; this is equivalent to measuring entropy vs energy \cite{mora+bialek_11,stephens+al_13}.  Strikingly, for the activity of neurons in the retina, the entropy is essentially a linear function of the energy, with unit slope \cite{tkacik+al_15b}, which corresponds to an unusual kind of critical point.    

There is an independent literature that tries to connect the dynamical patterns of activity in neural systems with the scale--invariant ``avalanches'' predicted by self--organized criticality \cite{beggs+plenz_03,beggs+plenz_04,friedman+al_12}.  Another dynamical notion of criticality is to ask about the number of Lyapunov exponents near zero, and there is an elegantly simple model  that shows how a network could learn to be critical in this sense \cite{magnasco+al_09}.  Subsequent work from Magnasco and colleagues has looked at the data emerging from human electro--corticography;  they estimate  the spectra of Lyapunov exponents for models that describe these dynamical signals, show that there is  a concentration of exponents near zero, and even  that this critical behavior is lost as the patient slips out of consciousness under anesthesia \cite{solovey+al_12,solovey+al_15}.  The relationship between statistical and dynamical notions of criticality is not at all clear, and this is a physics problem not a biology problem; for a first try at connecting the different ideas in the context of neural data, see Ref \cite{mora+al_15a}.

Returning to the families of proteins, we again see hints of critical behavior.  The hope is that the distribution of sequences can be described by models in which the different choices of amino acid interact only when the residues are in contact, but we also know that measured correlations extend over long distances, which is why the attempt to infer contacts from correlations is hard.  If this picture really is correct, we have the coexistence of local interactions and long--ranged correlations, which is a signature of criticality.  But the situation is far from clear, since the data  are still sparse,\footnote{For related reasons, I have not discussed the problem of genetic networks, although it does seem appropriate to give some pointers.  There is early work connecting biochemical and genetic networks to Boolean networks \cite{kauffman_69}, and this led to substantial theoretical developments \cite{derrida+flyvbjerg_86,derrida+pomeau_86}.  A second wave used ideas borrowed from Hopfield's approach to neural networks \cite{mjolsness+al_91}, while more recent work has focused on incorporating what we know of the molecular details \cite{bintu+al_05a,bintu+al_05b}.  Picking up old threads, Kauffman and colleagues made efforts to identify signatures of criticality in genetic networks \cite{shmulevich+al_05,nykter+al_08,balleza+al_08}, while my colleagues and I have argued that one can see such signatures in the the statistical and dynamical behavior of the gap gene network \cite{krotov+al_14}.  I think all will be clearer when we finally have tools that allow us to measure simultaneously the expression levels of many genes, in single cells, with a resolution significantly better than the intrinsic noise levels, and these are just emerging \cite{lubeck+cai_12,chen+al_15}.}    and correlations derived from functionality are mixed with correlations derived from shared evolutionary history.  We have tried a test case---the diversity of antibodies in the zebrafish immune system---that involves much shorter sequences,  where the relevant protein family can be exhaustively sampled \cite{weinstein+al_09}, and hence where the  maximum entropy construction can be carried, convincingly, to completion.  Even in this more limited problem, we see signs that the distribution of sequences is poised near a critical point in parameter space \cite{mora+al_10}.

\section{Toward conclusions}

I hope to have convinced you that our modern understanding of the phenomena of life has already been influenced, dramatically, by theory, and that the prospects for the future are bright.   This is, perhaps, a moment to emphasize that the examples I have chosen are far from exhaustive.  In the same spirit, I could  have discussed many other beautiful developments:  the idea that reliable transmission of information through the synthesis of new molecules---as in the replication, transcription, and translation of sequence information coded in DNA---depends on building Maxwell demons (kinetic proofreading) that can push past the limits to precision set by thermodynamics \cite{hopfield_74,ninio_75,hopfield_80};  the idea that amino acid sequences of real proteins are selected to avoid the frustration that leads to the glassiness of random heteropolymers \cite{bryngelson+wolynes_87,leopold+al_92,onuchic+al_95};  the idea that the pace of evolutionary change is determined not by the typical organism, but by those rare organisms in the tail of the fitness distribution, as well as broader connections of evolutionary dynamics to statistical physics \cite{desai+al_07,neher+al_10,hallatschek_11,neher+shraiman_11,fisher_12}; the idea that the active mechanics of the inner ear are tuned near a critical point (Hopf bifurcation), maximizing sensitivity and frequency selectivity while providing a natural and nearly parameter--free explanation for the essential nonlinearities of auditory perception \cite{eguiluz+al_00,calamet+al_00,magnasco_03};  and more.\footnote{Beyond these theoretical ideas, there are examples of striking phenomenology that still haven't been given the full theoretical treatment that they deserve.  At the top of this list, for me, is the observation of quantum coherence in the early events of photosynthesis \cite{vos+al_94,engel+al_07}, which should provide a stimulus to examine, more generally, our understanding of the quantum/classical boundary in the dynamics of biological molecules.}

Despite these many examples, there is a persistent notion that biology has developed without significant theoretical input.  This is reinforced by what amounts to revisionist history in the teaching of biology.  If biology is presented to undergraduate students as the science they can do even if they don't like math, then when it comes time to teach them about the foundations of molecular and cellular neuroscience, one simply cannot write down the Hodgkin--Huxley equations and expect the students to understand what is going on.  Similarly, now that we can sequence DNA, it is conventional to  suppress the fact that the linear arrangement of genes along chromosomes was established by mathematical analysis, long before we even knew the identity of DNA as the molecule that carries genetic information \cite{sturtevant_13}.    Even when it comes to experimental methods, few modern biology curricula teach the theory of X--ray diffraction from a helix \cite{cochran+al}, and thus students do not learn the mathematics behind the interpretation of Rosalind Franklin's famous observations on DNA \cite{franklin+gosling_53}.  The message, I think, is that mathematical analysis---not to speak of theory---is merely technical.   Even with the proliferation of  graduate programs in quantitative biology, so long as this anti--mathematical  approach constitutes the mainstream of biology teaching, we cannot expect that the biology community itself will create a genuinely receptive audience for theory.  If the community insists that what is ``biologically relevant'' must always be translated into words, then the search for mathematical description can never be central to the practice of biology.    In a dissent from cheerful interdisciplinarity, I believe it is essential that the physics community   provide a home for the theoretical physics of biological systems.

Discussions of the relation between physics and biology, and especially of the relation between theoretical physics and biology, often include various warnings about theorists isolating themselves from experiment, running off to do things which are irrelevant.    I believe that these concerns are wildly overstated.  My colleagues and I, who are trying to do theory at the interface of physics and biology, spend quite a lot of our time interacting with experiments, and with experimentalists.  Indeed, one of the traditional roles of theory in physics is to highlight things that would be interesting to measure, and this happens as we try to theorize about biological systems as well.  Although it often is claimed that biology is awash in data, in fact the attempt to build theories often points to numbers that we don't know, numbers that can determine which of several theoretical directions is most productive.  Sometimes measuring these quantities that are most relevant for theory drives the development of new experimental methods, or new data analysis strategies, and these have implications well beyond the original theoretical ideas.\footnote{An example  is the idea that one can decode complex, dynamic sensory signals from sequences of action potentials \cite{spikes}.  This ``stimulus reconstruction'' grew out our interest   in measuring the reliability and precision of neural computation under conditions closer to those found in  nature (\S\ref{info}).  The theoretical work involved both understanding the physical limits to reliability (see above) and developing conditions under which decoding could be simple even when encoding was complicated.  While I still think the results on the precision of computation are very important, the idea of decoding itself had a much larger impact, and even had implications for practical matters such as neural prostheses.}  This means, in particular, that theories can be enormously productive even if they are wrong, or not faithful to all the details of the real systems we are thinking about.

If you are worried about a disconnect between theory and experiment, I think that there is a much greater danger of people doing experiments and collecting data that will never fit into any mathematical framework.  This seems especially likely at a moment when you can collect exponentially more data than you could before.   I would remind you that in other data intensive, phenomenological areas---astrophysics and cosmology, for example---when you go off to spend $\sim$\$100 million to collect data, there are theorists on the team for the design of the instruments and observations.  You think about what you're looking for and what framework you're planning on analyzing it with {\em before} you collect the data, not after. 

The attentiveness of theorists to experiment also raises the worry that we will lose sight of our more grand ambitions.   It certainly is true that we live in an era where data is expanding exponentially, and this is a good thing. And we as theorists are the richer for it.   But theory is more than data mining. The point here is that miners know gold when they see it.  What you do when you are data mining is to look for certain kinds of structure;  within the set of possible structures you identify the one which is best supported by the data, and then pin down the parameters within  this best structure.  But the possible structures are, in a very real sense, your theories about what might be going on.   If your list of structures is not rich enough and deep enough, if your list of possible theories doesn't include the right one, you're not going to understand what's going on, and no amount of data is going to solve this problem.

Finally, I believe that  the deepest theoretical questions transcend the boundaries between the subfields of biology.  I hope that this is clear from the examples that I have given.   I am excited to see the same theoretical questions being formulated in  different biological contexts, in some cases really using the same mathematics to describe these very different systems.  One of the ways in which this has happened is by focusing on problems that the organism itself has to solve, from digging weak signals out of a noisy background to setting the parameters of its own networks.    Even if the answers are different, it is attractive to think of mechanisms in different systems, even at different levels of organization, as being chosen by Nature to solve the same physics problems that the organism faces in different contexts. Similarly, in the discussion of collective behavior, we have seen the same conceptual principles organizing our thinking about problems ranging from the evolution of protein families to the dynamics of flocks and swarms, not just at an abstract level but also engaging with details of the data.  Importantly, we see all these commonalities only through theory, and thus theory has the chance of redrawing the intellectual  landscape of  the field.

\section{Coda}

In looking more carefully through the references,  I realized that the spirit of what I want to convey here was expressed long ago, albeit in a different context \cite{crick_58}:

``What are one's overall impressions of the present
state of the subject? Two things strike me particularly. First, the existence
of general ideas covering wide aspects of the problem. It is remarkable
that one can formulate principles ... which explain many striking facts and yet for which proof
is completely lacking. This gap between theory and experiment is a great
stimulus to the imagination. Second, the extremely active state of the
subject experimentally ... new and significant results are being reported every few
months, and there seems to be no sign of work coming to a standstill
because experimental techniques are inadequate.''

What Crick was saying about the interplay between theory and experiment in the exploration of  the genetic code, now nearly sixty years ago, is something that applies today to our exploration of life much more broadly.

\begin{acknowledgments}
Thanks to the Simons Foundation, and to many colleagues involved in the 2014 workshop, for the opportunity to sharpen the ideas expressed here.  Thanks also to Lee Morgan for transcribing the lecture.  My own work on these problems has been in collaboration with many others, as can be seen from the reference list, who have made these explorations a pleasure.  We have been supported in part by the National Science Foundation, most recently through grants PHY--1305525, PHY--1451171, and CCF--0939370, by the Simons Foundation, and by the Swartz Foundation.  Special thanks to CG Callan and MO Magnasco, for many long conversations about what it is we all are trying to do. 
\end{acknowledgments}


\begin{thebibliography}{99}
%
\bibitem{bialek_12}
W Bialek, {\em Biophysics: Searching for Principles} (Princeton University Press, Princeton, 2012).
%
\bibitem{rayleigh_07}
Lord Rayleigh, XII. On our perception of sound direction.  {\em Phil Mag Series 6} {\bf 13,} 214--232 (1907).
%
\bibitem{watson+crick_53a}  
JD Watson and FHC Crick,  A structure for deoxyribose nucleic acid.  {\em Nature} {\bf 171,} 737--739 (1953).
%
\bibitem{watson+crick_53b}  
JD Watson and FHC Crick, Genetical implications of the structure of deoxyribonucleic acid.  {\em Nature} {\bf 171,} 964--967 (1953).
%
\bibitem{crick_58}
FHC Crick, On protein synthesis.  {\em Symp Soc Exp Biol } {\bf 12,} 138--163 (1958).
%
\bibitem{judson_79}
HF Judson, {\em The Eighth Day of Creation}. (Simon and Schuster, New York, 1979).
%
\bibitem{watson_80}
JD Watson, {\em The Double Helix: A Personal Account of the Discovery of the Structure of DNA}. Norton Critical Edition, G Stent, ed (Norton, New York, 1980).
%
\bibitem{maddox_02}
B Maddox, {\em Rosalind Franklin:  The Dark Lady of DNA}. (Harper Collins, 2002).
%
\bibitem{franklin+gosling_53}
RE Franklin and RG Gosling,  Molecular configuration in sodium thymonucleate. {\em Nature} {\bf  171,} 740--741 (1953).
%
\bibitem{wilkins+al_53}
MHF Wilkins, AR Stokes, and HR Wilson, Molecular structure of deoxypentose nucleic acids. {\em Nature} {\bf  171,} 738--740 (1953).
%
\bibitem{franklin+gosling_53b}
RE Franklin and RG Gosling, Evidence for 2--chain helix in crystalline structure of sodium deoxyribonucleate.  {\em Nature} {\bf 172,} 156--157 (1953).
%
\bibitem{franklin+gosling_53c}
RE Franklin and RG Gosling,  The structure of sodium thymonucleate fibres. I. The influence of water content. {\em Acta Cryst} {\bf 6,} 673--677 (1953).
%
\bibitem{franklin+gosling_53d}
RE Franklin and RG Gosling,  The structure of sodium thymonucleate fibres. II. The cylindrically symmetrical Patterson function. {\em Acta Cryst} {\bf 6,} 678--685 (1953).
%
\bibitem{langridge+al_60a}
R Langridge, HR Wilson, CW Hooper, MHF Wilkins, and LD Hamilton, The molecular configuration of deoxyribonucleic acid.  I. X--ray diffraction study of a crystalline form of the lithium salt.  {\em J Mol Biol} {\bf 2,} 19--37 (1960).
%
\bibitem{langridge+al_60b}
R Langridge, DA Marvin, WE Seeds, HR Wilson, CW Hooper, and MHF Wilkins, The molecular configuration of deoxyribonucleic acid.  II. Molecular models and their Fourier transforms.  {\em J Mol Biol} {\bf 2,} 38--64 (1960).
%
\bibitem{cochran+al}
W Cochran, FHC Crick, and V Vand,  The structure of synthetic polypeptides. I. The transform of atoms on a helix. {\em Acta Cryst} {\bf 5,} 581--586 (1952).
%
\bibitem{anderson_72}
PW Anderson, More is different.  {\em Science} {\bf 177,} 393--396 (1972). 
%
\bibitem{hodgkin+huxley_52}
AL Hodgkin and AF Huxley, A quantitative description of membrane current and its application
to conduction and excitation in nerve.  {\em J Physiol (Lond)} {\bf 117,}
500--544 (1952).
%
\bibitem{spikes}
F Rieke, D Warland, R de Ruyter van Steven\-inck, and W Bialek,  {\em Spikes: Exploring the Neural Code.}  (MIT Press, Cambridge, 1997).
%
\bibitem{lemasson+al_93}
G LeMasson, E Marder, and LF Abbott, Activity--dependent regulation of conductances in model neurons.  {\em Science} {\bf 259,} 1915--1917 (1993).
%
\bibitem{abbott+lemasson_93}
LF Abbott and G LeMasson, Analysis of neuron modelswith dynamically regulated conductances.  {\em Neural Comp} {\bf 5,} 823--842 (1993).
%
\bibitem{goldman+al_01}
MS Goldman, J Golowasch, E Marder, and LF Abbott, Global structure, robustness, and modulation of neuronal models. {\em J Neurosci} {\bf21,} 5229--5238 (2001).
%
\bibitem{liu+al_98}
Z Liu, J Golowasch, E Marder, and LF Abbott, A model neuron with activity--dependent conductances regulated by multiple calcium sensors.  {\em J Neurosci} {\bf 18,} 2309--2320
(1998).
%
\bibitem{prinz+al_04}
AA Prinz, D Bucher, and E Marder,  Similar network activity from disparate circuit parameters.  {\em Nat Neurosci} {\bf 7,} 1345--1352 (2004).
%
\bibitem{turrigiano+al_94}
G Turrigiano, LF Abbott, and E Marder,  Activity--dependent changes in the intrinsic properties of cultured neurons. {\em Science} {\bf 264,} 974--977 (1994).
%
\bibitem{cahn_96}
RN Cahn, The eighteen parameters of the standard model in your everyday life.  {\em Rev Mod Phys} {\bf 68,} 952--959 (1996).
%
\bibitem{bcs}
J Bardeen, LN Cooper, and JR Schrieffer, Theory of superconductivity.  {\em Phys Rev} {\bf 108,} 1175--1204 (1957).
%
\bibitem{wilson_75}
KG Wilson, The renormalization group:  Critical phenomena and the Kondo problem.  {\em Rev Mod Phys} {\bf 47,} 773--840 (1975).
%
\bibitem{laughlin_83}
RB Laughlin, Anomalous quantum Hall effect: An incompressible quantum fluid with fractionally charged excitations.  {\em Phys Rev Lett} {\bf 50,} 1395--1398 (1983).
%
\bibitem{gross+wilczek_74}
DJ Gross and F Wilczek, Asymptotically free gauge theories. II.  {\em Phys Rev D} {\bf 9,} 980--993 (1974).
%
\bibitem{axion}
F Wilczek, Problems of strong P and T invariance in the presence of instantons.  {\em  Phys Rev Lett} {\bf 40,} 279--282 (1977).
%
\bibitem{barkai+leibler_97}
N Barkai and S Leibler, Robustness in simple biochemical networks. 
{\em Nature} {\bf 387,} 913--917 (1997).
%
\bibitem{dassow+al_00}
 G von Dassow, E Meir, EM Munro, and GM Odell,  The segment polarity network is a robust developmental module. {\em Nature} {\bf 406,} 188--192 (2000).
 %
 \bibitem{gregor+al_07}
 T Gregor, DW Tank, EF Wieschaus, and W Bialek, Probing the limits to positional information.  {\em Cell} {\bf 130,} 153--164 (2007).
 %
 \bibitem{petkova+al_14}
 MD Petkova, SC Little, F Liu, and T Gregor,  Maternal origins of developmental reproducibility. {\em Curr Biol} {\bf 24,} 1283--1288 (2014).
 %
\bibitem{kollmann+al_05}
M Kollmann, L L{\o}vdok, K Bartholome, J Timmer, and V Sourjik   Design principles of a bacterial signalling network. {\em Nature} {\bf 438,} 504--507 (2005).
 %
 \bibitem{lovdok+al_09}
L L{\o}vdok, K Bentele, N Vladimirov,  A M\"uller, FS Pop, D Lebiedz, M Kollmann, and V Sourjik,  Role of translational coupling in robustness of bacterial chemotaxis pathway.
{\em PLoS Biology} {\bf 7,} e1000171 (2009).
%
\bibitem{hopfield_82}
JJ Hopfield,  Neural networks and physical systems with emergent collective computational abilities. {\em Proc Natl Acad Sci USA} {\bf 79,} 2554--2558 (1982).
%
\bibitem{seung_96}
HS Seung, How the brain keeps the eyes still.   {\em Proc Natl Acad Sci (USA)} {\bf 93,} 13339--13334 (1996).
%
\bibitem{major+al_04a}
G Major, R Baker, E Aksay, B Mensh, HS Seung, and DW Tank, Plasticity and tuning by visual feedback of the stability of a neural integrator.
 {\em Proc Natl Acad Sci (USA)} {\bf 101,} 7739--7744 (2004).
%
\bibitem{major+al_04b}
G Major, R Baker, E Aksay, HS Seung, and DW Tank,  Plasticity and tuning of the time course of analog persistent firing in a neural integrator.  {\em Proc Natl Acad Sci (USA)} {\bf  101,} 7745--7750 (2004).
%
\bibitem{ruyter+al_86}
RR de Ruyter van Steveninck, WH Zaagman, and HAK Mastebroek, Adaptation of transient responses of a movement--sensitive neuron in the visual system of the blowfly {\em Calliphora erythrocephala}. {\em Biol Cybern} {\bf 54,}  223--236 (1986).
%
\bibitem{ruyter+laughlin_96a}
RR de Ruyter van Steveninck and SB Laughlin,  Light adaptation and reliability in blowfly
photoreceptors. R de Ruyter van Steveninck and SB Laughlin, {\em Int J Neural Syst} {\bf 7,} 437--444 (1996).
%
\bibitem{ruyter+laughlin_96b}
RR de Ruyter van Steveninck and SB Laughlin, The rate of information transfer at graded potential synapses. {\em Nature} {\bf 379,} 642--645 (1996).
%
\bibitem{bialek+al_91}
W Bialek, F Rieke, RR de Ruyter van Ste\-ven\-inck, and D Warland, Reading a neural code.  {\em Science} {\bf 252,} 1854--1857 (1991).
%
\bibitem{ruyter+bialek_95}
R de Ruyter van Steveninck and W Bialek,  Reliability and statistical efficiency of a blowfly movement--sensitive neuron.  {\em Phil Trans R. Soc Lond.} {\bf 348,} 321--340 (1995).
%
\bibitem{berg+purcell_77}
HC Berg and EM Purcell, Physics of chemoreception. {\em Biophys J} {\bf 20,} 193--219 (1977).
%
\bibitem{bialek+setayeshgar_05}
W Bialek and S Setayeshgar,  Physical limits to biochemical signaling.  {\em Proc Natl Acad Sci (USA)} {\bf 102,} 10040--10045 (2005).
%
\bibitem{bialek+setayeshgar_08}
W Bialek and S Setayeshgar,  Cooperativity, sensitivity and noise in biochemical signaling.   {\em Phys Rev Lett} {\bf 100,} 258101 (2008).
%
\bibitem{tkacik+bialek_09}
G Tka\v{c}ik and W Bialek, Diffusion, dimensionality and noise in transcriptional regulation.  {\em Phys Rev E} {\bf 79,} 051901 (2009).
%
\bibitem{endres+wingreen_09}
RG Endres and NS Wingreen, Maximum likelihood and the single receptor. {\em Phys Rev Lett} {\bf 103,}  158101 (2009).
%
\bibitem{mora+wingreen_10}
T Mora and NS Wingreen, Limits of sensing temporal concentration changes by single cells.  {\em Phys Rev Lett} {\bf 104,} 248101 (2010).
%
\bibitem{govern+wolde_12}
CC Govern and PR ten Wolde, Fundamental limits on sensing chemical concentrations with linear biochemical networks. {\em Phys Rev Lett} {\bf 109,} 218103 (2012).
%
\bibitem{kaizu+al_14}
K Kaizu, WH de Ronde, J Paijmans, K Takahashi, F Tostevin, and PR ten Wolde, The Berg--Purcell limit revisited. {\em Biophys J} {\bf 106,} 976--985 (2014).
%
\bibitem{paijmans+al_14}
J Paijmans and PR ten Wolde, Lower bound on the precision of transcriptional regulation and why facilitated diffusion can reduce noise in gene expression. {\em Phys Rev E} {\bf 90,}032708 (2014).
%
\bibitem{tkacik+al_08}
G Tka\v{c}ik, T Gregor, and W Bialek,   The role of input noise in transcriptional regulation.   {\em PLoS One} {\bf 3,} e2774 (2008).
%
\bibitem{potters+bialek_94}
M Potters and  W Bialek,  Statistical mechanics and visual signal processing. {\em J Phys I France} {\bf 4}, 1755--1775 (1994).
%
\bibitem{ruyter+al_94}
RR de Ruyter van Steveninck, W Bialek, M Potters, and RH Carlson,
Statistical adaptation and optimal estimation in movement
computation by the blow\-fly visual system.  In {\em Proc IEEE Conf Sys Man Cybern}, 302--307 (1994).
%
\bibitem{bialek+ruyter_05}
W Bialek and R de Ruyter van Steveninck, Features and dimensions: Motion estimation in fly vision.  arXiv:q--bio/0505003 (2005).
%
\bibitem{bialek+owen_90}
W Bialek and WG Owen, Temporal filtering in retinal bipolar cells: Elements of an optimal computation? {\em Biophys J} {\bf 58,} 1227--1233 (1990).
%
\bibitem{rieke+al_91}
F Rieke, WG Owen, and W Bialek, Optimal filtering in the salamander retina.
In {\em Advances in Neural Information Processing 3,} R Lippman, J Moody \& D Tour\-etzky, eds, pp 377--383 (Morgan Kaufmann, San Mateo CA, 1991).
%
\bibitem{field+rieke_02}
GD Field and F Rieke, Nonlinear signal transfer from mouse rods to bipolar cells and implications for visual sensitivity.  {\em Neuron} {\bf 34,} 773--785 (2002).
%
\bibitem{weiss+al_02}
Y Weiss, EP Simoncelli, and EH Adelson, Motion illusions as optimal percepts. {\em Nat Neurosci} {\bf 5,} 598--604 (2002).
%
\bibitem{stocker+simoncelli_06}
AA Stocker and EP Simoncelli,   Noise characteristics and prior expectations in human visual speed perception. {\em Nat Neurosci} {\bf 9,} 578--585 (2006).
%
\bibitem{fitzgerald+al_11}
JE Fitzgerald, AY Katsov, TR Clandinin, and MJ  Schnitzer, Symmetries in stimulus statistics shape the form of visual motion estimators. {\em Proc Natl Acad Sci
(USA)} {\bf 108,} 12909--12914 (2011).
%
\bibitem{clark+al_14}
DA Clark,  JE Fitzgerald, JM Ales, DM Gohl, MA Silies, AM Norcia, and TR Clandinin,  Flies and humans share a motion estimation strategy that exploits natural scene statistics. {\em Nat Neurosci} {\bf 17,} 296--303 (2014).
%
\bibitem{fitzgerald+clark_15}
JE Fitzgerald and DA Clark, Nonlinear circuits for naturalistic visual motion
estimation. {\em eLife}  {\bf 4,} e09123 (2015).
%
\bibitem{takemura+al_13} S Takemura, et al,  A visual motion detection circuit suggested by {\em Drosophila} connectomics.  {\em Nature} {\bf 500,} 175--181 (2013).
%
\bibitem{fisher+al_15}
YE Fisher, JCS Leong, K Sporar, MD Ketkar, DM Gohl, TR Clandinin, and M Silies, A class of visual neurons with wide--field properties is required for local motion detection. {\em Curr Biol} {\bf 25,} 3178--3189 (2015).
%
 \bibitem{attneave_54}
F Attneave,  Some informational aspects of visual perception.   {\em Psych Rev} {\bf 61,} 183--193 (1954).
%
 \bibitem{barlow_59}
HB Barlow, Sensory mechanisms, the reduction of redundancy, and intelligence.  In 
 {\em  Proceedings of the Symposium on the Mechanization of Thought Processes, volume 2}, DV Blake and AM Utlley, eds, pp 537--574 (HM Stationery Office, London, 1959).
%
\bibitem{barlow_61}
HB Barlow, Possible principles underlying the transformation of sensory messages.  In  {\em Sensory Communication}, W Rosenblith, ed, pp 217--234 (MIT Press, Cambridge, 1961).
%
\bibitem{mackay+mcculloch_52}
D MacKay and WS McCulloch, The limiting information capacity of a neuronal link.  {\em Bull Math Biophys} {\bf 14,} 127--135 (1952).
%
\bibitem{rieke+al_93}
 F Rieke, D Warland, and  W Bialek, Coding efficiency and information rates in sensory neurons.  {\em Europhys. Lett} {\bf 22,} 151--156 (1993).
 %
 \bibitem{strong+al_98a}
SP Strong, R Koberle, RR de Ruyter van Steveninck, and W Bialek,  Entropy and information in neural spike trains.  {\em Phys Rev Lett} {\bf 80,} 197--200 (1998).
%
 \bibitem{strong+al_98b}
SP Strong, RR de Ruyter van Steveninck, W Bialek, and R Koberle, On the application of information theory to neural spike trains.  In {\em Pacific Symposium on Biocomputing `98}, RB Altman, AK Dunker, L Hunter, and TE Klein, eds, pp 621--632 (World Scientific, Singapore, 1998).
%
\bibitem{rieke+al_95}
F Rieke, DA Bodnar, and W Bialek,  Naturalistic stimuli increase the rate and efficiency of information transmission by primary auditory neurons.  {\em Proc R Soc Lond Ser. B} {\bf 262,} 259--265 (1995).
%
\bibitem{lewen+al_01}
GD Lewen, W Bialek, and RR de Ruyter van Steveninck,  Neural coding of naturalistic motion stimuli.  {\em Network} {\bf 12,} 317--329 (2001).
%
\bibitem{wright+al_02}
BD Wright, K Sen, W Bialek, and AJ Doupe,  Spike timing and the coding of naturalistic sounds in a central auditory area of songbirds.  In {\em Advances in Neural Information Processing 14,}  TG Dietterich, S Becker, and Z Ghahramani, eds, pp 309--316 (MIT Press, Cambridge, 2002).
%
\bibitem{nemenman+al_08}
GD Lewen, W Bialek, and RR de Ruyter van Steveninck,  Neural coding of a natural stimulus ensemble:  Information at sub--millisecond resolution.  I Nemenman, {\em PLoS Comput Biol} {\bf 4,} e1000025 (2008).
%
\bibitem{laughlin_81}
SB Laughlin, A simple coding procedure enhances a neuron's information capacity. {\em Z Naturforsch} {\bf 36c,} 910--912 (1981).
%
\bibitem{ruderman+bialek_94}
DL Ruderman and W Bialek, Statistics of natural images: Scaling in the woods.   {\em Phys Rev Lett} {\bf 73}, 814--817 (1994).
%
\bibitem{smirnakis+al_97}
S Smirnakis, MJ Berry II, DK Warland, W Bialek, and M Meister,  Adaptation of retinal processing to image contrast and spatial scale. {\em Nature} {\bf 386,} 69--73 (1997).
%
\bibitem{brenner+al_00a}
N Brenner, W Bialek, and R de Ruyter van Steveninck, Adaptive rescaling optimizes information transmission. {\em Neuron} {\bf 26,} 695--702 (2000).
%
\bibitem{kvale+schreiner_04}
MN Kvale and CE Schreiner, Short-term adaptation of auditory receptive fields to dynamic stimuli. {\em J Neurophysiol} {\bf 91,} 604--612 (2004).
%
\bibitem{dean+al_05}
I Dean, NS Harper, and D McAlpine, Neural population coding of sound level adapts to stimulus statistics.  {\em Nature Neurosci} {\bf 8,} 1684--1689 (2005).
%
\bibitem{nagel+doupe_06}
KI Nagel and AJ Doupe, Temporal processing and adaptation in the songbird auditory forebrain.  {\em Neuron} {\bf 21,} 845--859 (2006).
%
\bibitem{maravall+al_07}
M Maravall, RS Petersen, AL Fairhall, E Arabzadeh, and ME Diamond, Shifts in coding properties and maintenance of information transmission during adaptation in barrel cortex. {\em PLoS Biology} {\bf 5,} e19 (2007).
%
\bibitem{debaene+al_07}
W De Baene, E Premereur, and R Vogels, Properties of shape tuning of macaque inferior temporal neurons examined using rapid serial visual presentation.  {\em J Neurophysiol} {\bf 97,} 2900--2916 (2007).
%
\bibitem{wen+al_09}
B Wen, GI Wang, I Dean, and B Delgutte, Dynamic range adaptation to sound level statistics in the auditory nerve.  {\em J Neurosci} {\bf 29,} 13797--13808 (2009).
%
\bibitem{dahmen+al_10}
JC Rahmen, P Keating, FR Nodal, AL Schulz, and AJ King, Adaptation to stimulus statistics in the perception and neural representation of auditory space. {\em Neuron} {\bf 66,} 937--948 (2010).
%
\bibitem{rabinowitz+al_11}
NC Rabinowitz, BDB Willmore, JWH Schnup, and AJ King, Contrast gain control in auditory cortex.  {\em Neuron} {\bf 70,} 1178--1192 (2011).
 %
\bibitem{fairhall+al_01}
AL Fairhall, GD Lewen, W Bialek, and RR de Ruyter van  Steveninck,   Efficiency and ambiguity in an adaptive neural code. {\em Nature}  {\bf 412,} 787--792  (2001).
%
\bibitem{wark+al_09}
B Wark, A Fairhall, and F Rieke, Timescales of inference in visual adaptation.  {\em Neuron} {\bf 61,} 750--761 (2009).
%
\bibitem{tkacik+al_08a}
G Tka\v{c}ik, CG Callan Jr, and W Bialek, Information capacity of genetic regulatory elements.  {\em Phys Rev E} {\bf 78,} 011910 (2008).
%
\bibitem{tkacik+al_08b}
G Tka\v{c}ik, CG Callan Jr, and W Bialek,  Information flow and optimization in transcriptional regulation.  {\em Proc Natl Acad Sci (USA)} {\bf 105,} 12265--12270 (2008).
%
\bibitem{dubuis+al_13}
 JO Dubuis, G Tka\v{c}ik, EF Wieschaus, T Gregor, and W Bialek,  Positional information, in  bits. {\em Proc Natl Acad Sci (USA)} {\bf 110,}  16301--16308 (2013).
 %
\bibitem{tkacik+al_09a}
G Tka\v{c}ik, AM Walczak, and W Bialek,   Optimizing information flow in small genetic networks.   {\em Phys Rev E} {\bf 80,} 031920 (2009).
%
\bibitem{walczak+al_10}
AM Walczak, G Tka\v{c}ik, and W Bialek, Optimizing information flow in small genetic networks. II: Feed--forward interaction.   {\em Phys Rev E} {\bf 81,} 041905 (2010).
%
\bibitem{tkacik+al_12}
G Tka\v{c}ik, AM Walczak, and  W Bialek,  Optimizing information flow in small genetic networks. III. A self--interacting gene.   {\em Phys Rev E} {\bf 85,}  041903 (2012).
%
\bibitem{sokolowski+al_15a}
TR Sokolowski  and G Tka\v{c}ik,  Optimizing information flow in small genetic networks. IV. Spatial coupling.  {\em Phys Rev E} {\bf 91,} 062710 (2015).
%
\bibitem{sokolowski+al_15b}
TR Sokolowski, AM Walczak, W Bialek, and G Tka\v{c}ik,  Extending the dynamic range of transcription factor action by translational regulation. arXiv.org:1507.02562 [q--bio.MN] (2015).
%
\bibitem{bialek+al_01}
W Bialek, I Nemenman, and N Tishby,  Predictability, complexity and learning. {\em Neural Comp} {\bf 13,} 2409--2463  (2001).
 %
 \bibitem{bialek+al_07}
 W Bialek, RR de Ruyter van Steveninck, and N Tishby, Efficient representation as a design principle for neural coding and computation.
 arXiv:0712.4381 [q--bio.NC] (2007). A preliminary account appears in the {\em Proceedings of the International Symposium on Information Theory 2006}.
%
\bibitem{palmer+al_15}
SE Palmer, O Marre, MJ Berry II, and W Bialek,  Predictive information in a sensory population.   {\em Proc Natl Acad Sci (USA)} {\bf 112,} 6908--6913 (2015).
%
\bibitem{segev+al_04}
R Segev, J Goodhouse, J Puchalla, and  MJ Berry II,  Recording spikes from a large fraction of the ganglion cells in a retinal patch.  {\em Nature Neurosci} {\bf 7,} 1155--1162 (2004).
%
\bibitem{marre+al_12}
O Marre, D Amodei, K Sadeghi, F Soo, TE Holy, and MJ Berry II,   Recording from a complete population in the retina. {\em J Neurosci} {\bf 32,} 14859--14873  (2012).
%
\bibitem{schneidman+al_06}
E Schneidman, MJ Berry II, R Segev, and W Bialek, Weak pairwise correlations imply strongly correlated network states in a neural population.   {\em Nature} {\bf 440,} 1007--1012 (2006).
%
\bibitem{tkacik+al_14a}
G Tka\v{c}ik, O Marre, D Amodei, E Schneidman, W Bialek, and MJ Berry II,  Searching for collective behavior in a large network of sensory neurons.  {\em PLoS Comput Biol} {\bf 10,} e1003408 (2014).
%
\bibitem{tkacik+al_06}
G Tka\v{c}ik, E Schneidman, MJ Berry II, and W Bialek,  Ising models for networks of real neurons.  arXiv:q--bio.NC/0611072 (2006).
%
\bibitem{tkacik+al_09}
G Tka\v{c}ik, E Schneidman, MJ Berry II, and W Bialek,  Spin glass models for networks of real neurons.   arXiv:0912.5409 [q--bio.NC] (2009).
%
\bibitem{tkacik+al_12a}
G Tka\v{c}ik, O Marre, D Amodei, MJ Berry II, and W Bialek,  The simplest maximum entropy model for collective behavior in a neural network.  {\em J Stat Mech} P03011 (2013).
%
\bibitem{meshulam+al_15}
L Meshulam, J Gauthier, DW Tank, and W Bialek, Interpreting collective neural activity underlying spatial navigation in virtual reality.  {\em Bulletin of the APS}  {\bf 60(1),} G50.9 (2015).
%
\bibitem{socolich_al_05}
M Socolich, SW Lockless, WP Russ, H Lee, KH Gardner, and R Ranganathan, Evolutionary information for specifying a protein fold.   {\em Nature} {\bf 437,} 512--518 (2005).
%
\bibitem{russ+al_05}
WP Russ, DM Lowery, P Mishra, MB Yaffe, and R Ranganathan, Natural--like function in artificial WW domains.  {\em Nature} {\bf 437,} 579--583 (2005).
%
\bibitem{bialek+ranganathan_07}
W Bialek and R Ranganathan,  Rediscovering the power of pairwise interactions.  arXiv.org:0712.4397 [q--bio.QM] (2007).
%
%
\bibitem{lapedes+al_98}
AS Lapedes, BG Giraud, LC Liu, and GD Stormo,  A maximum entropy formalism for disentangling chains of correlated sequence positions.    In  {\em Proceedings of the IMS/AMS International Conference on Statistics in Molecular Biology and Genetics}   pp 236--256 (1998).
%
\bibitem{giraud+al_99}
BG Giraud, JM Heumann, and AS Lapedes, Superadditive correlation.  {\em Phys Rev E} {\bf 59,} 4983--4991 (1999).
%
\bibitem{lapedes+al_02}
A Lapedes, B Giraud, and C Jarzynski,   Using sequence alignments to predict protein structure and stability with high accuracy.   {\em Los Alamos National Laboratory Report} LA--UR--02--4481 (2002).  Later deposited at arXiv.org:1207.2484 [q--bio.QM] (2012).
%
\bibitem{weigt+al_09}
M Weigt, RA White, H Szurmant, JA Hoch, and T Hwa,  Identification of direct residue contacts in protein--protein interaction by message passing.  {\em Proc Natl Acad Sci (USA)} {\bf 106,} 67--72 (2009).
%
\bibitem{marks+al_11}
DS Marks, LJ Colwell, R Sheridan, TA Hopf, A Pagnani, R Zecchina, and C Sander, Protein 3D structure computed from evolutionary sequence variation.   {\em PLoS One} {\bf 6,} e28766 (2011). 
%
\bibitem{sulkowska+al_12}
 JI Sulkowska, F Morcos, M Weigt, T Hwa, and JN Onuchic,    Genomics--aided structure prediction.  {\em Proc Natl Acad Sci (USA)} {\bf 109,} 10340--10345 (2012). 
%
\bibitem{toner+tu_95}
J Toner and Y Tu,  Long--range order in a two--dimensional XY model: How birds fly together.  {\em Phys Rev Lett} {\bf 75,} 4326--4329 (1995).
%
\bibitem{vicsek+al_95}
T Vicsek, A Czir\'ok, E Ben--Jacob, I Cohen, and O Shochet,  Novel type of phase transition in a system of self--driven particles.  {\em Phys Rev Lett} {\bf 75,} 1226--1229 (1995).
%
\bibitem{toner+tu_98}
J Toner and Y Tu,  Flocks, herds, and schools: A quantitative theory of flocking.  {\em Phys Rev E} {\bf 58,} 4828--4858 (1998).
%
\bibitem{ramaswamy_10}
S Ramaswamy, The mechanics and statistics of active matter. {\em Annu Rev Cond Matt Phys} {\bf 1,} 323--345 (2010).
%
\bibitem{ballerini+al_08a}
M Ballerini, N Cabibbo, R Candelier, A Cavagna, E Cisbani, I Giardina, A Orlandi, G Parisi, A Procaccini, M Viale, and V Zdravkovic,  Empirical investigation of starling flocks:  A benchmark study in collective animal behaviour.   {\em Animal Behaviour} {\bf 76,} 201--215 (2008).
 %
\bibitem{cavagna+al_08a}
   A Cavagna,  I Giardina, A Orlandi, G Parisi, A Procaccini, M Viale, and V Zdravkovic,
  The STARFLAG handbook on collective animal behaviour: 1. Empirical methods.
 {\em Animal Behaviour} {\bf 76,} 217--236 (2008).
 %
\bibitem{cavagna+al_08b}
A Cavagna,  I Giardina, A Orlandi, G Parisi, and  A Procaccini, 
 The STARFLAG handbook on collective animal behaviour: 2.  Three--dimensional analysis.   {\em Animal Behaviour} {\bf 76,} 237--248 (2008).
 %
\bibitem{ballerini+al_08b}
M Ballerini, N Cabibbo, R Candelier, A Cavagna, E Cisbani, I Giardina, V Lecomte, A Orlandi, G Parisi, A Procaccini, M Viale, and  V Zdravkovic,  Interaction ruling animal collective behavior depends on topological rather than metric distance: Evidence from a field study.    {\em Proc Natl Acad Sci (USA)} {\bf 105,} 1232--1237 (2008).
 %
\bibitem{bialek+al_12}
 W Bialek, A Cavagna, I Giardina, T Mora, E Silvestri, M Viale, and A Walczak,   Statistical mechanics for natural flocks of birds. {\em Proc Natl Acad Sci (USA)} {\bf 109,} 4786--4791 (2012).
%
\bibitem{bialek+al_14}
W Bialek, A Cavagna, I Giardina, T Mora, O Pohl, E Silvestri, M Viale, and  AM Walczak,  Social interactions dominate speed control in poising natural flocks near criticality.  {\em Proc Natl Acad Sci (USA)} {\bf 111,} 7212--7217 (2014).
%
 \bibitem{cavagna+al_10}
 A Cavagna, A Cimarelli, I Giardina, G Parisi, R Santigati, F Stefanin, and M Viale, Scale--free correlations in starling flocks.  {\em Proc Natl Acad Sci (USA)} {\bf 107,} 11865--11870 (2010).%
\bibitem{cavagna+al_15}
A Cavagna, L Del Castillo, S Dey, I Giardina, S Melillo, L Parisi, and M Viale, 
Short--range interaction vs long--range correlation in bird flocks.   {\em Phys Rev E} {\bf 92,} 012705 (2015).
%
\bibitem{cavagna+al_14}
A Cavagna, I Giardina, F Ginelli, T Mora, D Piovani, R Tavarone, and AM Walczak,  Dynamical maximum entropy approach to flocking.  {\em Phys Rev E} {\bf 89,} 042707 (2014).
%
\bibitem{mora+al_15b}
T Mora, AM Walczak, L Del Castello, F Ginelli, S Melillo, L Parisi, M Viale, A Cavagna, and I Giardina, Questioning the activity of active matter in natural flocks of birds.  arXiv:1511.01958 [q--bio.PE] (2015).
%
\bibitem{attanasi+al_14a}
A Attanasi, A Cavagna, L Del Castello, I Giardina,   S Melillo, L Parisi, O Pohl, B Rossaro, E Shen, E Silvestri, and M Viale,  Collective behavior without collective order in wild swarms of midges.   {\em PLoS Comput Biol} {\bf 10,} e1003697 (2014). 
%
\bibitem{attanasi+al_14b}
A Attanasi, A Cavagna, L Del Castello, I Giardina, S Melillo, L Parisi, O Pohl, B Rossaro, E Shen, E Silvestri, and M Viale,  Finite--size scaling as a way to probe near--criticality in natural swarms.  {\em Phys Rev Lett} {\bf 113,} 238102 (2014).
%
\bibitem{mora+bialek_11}
T Mora and W Bialek,  Are biological systems poised at criticality?   {\em J Stat Phys} {\bf 144,} 268--302 (2011).
%
\bibitem{stephens+al_13}
GJ  Stephens, T Mora, G Tka\v{c}ik, and W Bialek, Statistical thermodynamics of natural images.  {\em Phys Rev Lett} {\bf 110,} 018701 (2013).
%
\bibitem{tkacik+al_15b}
 G Tka\v{c}ik, T Mora, O Marre, D Amodei, SE Palmer, MJ Berry II, and W Bialek,   Thermodynamics for a network of neurons: Signatures of criticality.  {\em Proc Natl Acad Sci (USA)} {\bf 112,}  11508--11513 (2015).
 %
\bibitem{beggs+plenz_03}
JM Beggs and D Plenz,   Neuronal avalanches in neocortical circuits.   {\em J Neurosci} {\bf 23,} 167--177 (2003).
%
\bibitem{beggs+plenz_04}
JM Beggs and D Plenz,   Neuronal avalanches are diverse and precise patterns of activity that are stable for many hours in cortical slice cultures.   {\em J Neurosci} {\bf 24,} 5216--5229 (2004).
%
\bibitem{friedman+al_12}
N Friedman, S Ito, B Brinkman, M Shimono, RL DeVille, K Dahmen, JM Beggs, and TC Butler,  Universal critical dynamics in high resolution neuronal avalanche data.   {\em Phys Rev Lett} {\bf 108,} 208102  (2012).
%
\bibitem{magnasco+al_09}
MO Magnasco, O Piro, and GA Cecchi,  Self--tuned critical anti--Hebbian networks. 
 {\em Phys Rev Lett} {\bf 102,} 258102 (2009).
%
\bibitem{solovey+al_12}
G Solovey, KJ Miller, JG Ojemann, MO Magnasco, and GA Cecchi,   Self--regulated dynamical criticality in human ECoG.  {\em Front Integ Neurosci} {\bf 6,} 44 (2012).
%
\bibitem{solovey+al_15}
G Solovey, LM Alonso, T Yanagawa, N Fuji, MO Magnasco, GA Cecchi, and A Proekt,  Loss of consciousness is associated with stabilization of cortical activity.   {\em J Neurosci} {\bf 35,} 10866--10877 (2015).
%
\bibitem{mora+al_15a} 
T Mora, S Deny, and O Marre,   Dynamical criticality in the collective activity of a population of retinal neurons. {\em Phys Rev Lett} {\bf 115,} 078105 (2015).
%
\bibitem{kauffman_69}
SA Kauffman,   Metabolic stability and epigenesis in randomly constructed genetic nets.   {\em J Theor Biol} {\bf 22,} 437--467 (1969).
%
\bibitem{derrida+flyvbjerg_86}
B Derrida and H Flyvbjerg, Multivalley structure in Kauffman's model: analogy with spin glasses.   {\em J Phys A} {\bf 19,} L1003--1008 (1986).
%
\bibitem{derrida+pomeau_86}
B Derrida and Y Pomeau, Random networks of automata: a simple annealed approximation.   {\em Europhys Lett} {\bf 1,} 45--49 (1986).
%
\bibitem{mjolsness+al_91}
E Mjolsness, DH Sharp, and J Reinitz,   A connectionist model of development.  {\em J Theor Biol} {\bf 152,} 429--453 (1995).
%
\bibitem{bintu+al_05a}
L Bintu, NE Buchler,  HG Garcia, U Gerland, T Hwa, J Kondev, and R Phillips,  Transcriptional regulation by the numbers: models.     {\em Curr Opin Genet Dev} {\bf 15,} 116--124 (2005).
%
\bibitem{bintu+al_05b}
L Bintu, NE Buchler,  HG Garcia, U Gerland, T Hwa, J Kondev, T Kuhlman, and R Phillips,   Transcriptional regulation by the numbers: applications. {\em Curr Opin Genet Dev} {\bf 15,} 125--135 (2005).
%
\bibitem{shmulevich+al_05}
I Shmulevich, S Kauffman, and M  Aldana,  Eukaryotic cells are dynamically ordered or critical but not chaotic.  {\em Proc Natl Acad Sci (USA)} {\bf  102,} 13439--13444 (2005).
 %
\bibitem{nykter+al_08}
M Nykter, ND Price, M Aldana, SA Ramsey, SA Kauffman, LE Hood, O Yli--Harja, and I Shmulevich,  Gene expression dynamics in the macrophage exhibit criticality.  {\em Proc  Natl Acad  Sci  (USA)} {\bf 105,}  1897--1900 (2008). 
%
\bibitem{balleza+al_08}
E Balleza, ER Alvarez--Buylla, A Chaos, S Kauffman, I Shmulevich, and M Aldana,  Critical dynamics in genetic regulatory networks: Examples from four kingdoms.   {\em PLoS One} {\bf 3,} e 2456 (2008).
%
\bibitem{krotov+al_14}
D Krotov, JO Dubuis, T Gregor, and W Bialek,  Morphogenesis at criticality?  {\em Proc Natl Acad Sci (USA)} {\bf 111,} 3683--3688 (2014).
%
\bibitem{lubeck+cai_12}
E Lubeck and L Cai, Single--cell systems biology by super--resolution imaging and combinatorial labeling.   {\em Nature Meth} {\bf 9,} 743--748 (2012).
%
\bibitem{chen+al_15} 
KH Chen, AN Boettiger, JR Moffitt,  S Wang, and X Zhaung,   Spatially resolved, highly multiplexed RNA profiling in single cells. {\em Science} {\bf 348,} aaa6090 (2015).
%
\bibitem{weinstein+al_09}
JA Weinstein, N Jiang, RA White, DS Fisher, and SR Quake, High--throughput sequencing of the zebrafish antibody repertoire.  {\em Science} {\bf 324,} 807--810 (2009).
%
\bibitem{mora+al_10}
T Mora, AM Walczak, W Bialek, and CG Callan Jr,     Maximum entropy models for antibody diversity.   {\em Proc Natl Acad Sci (USA)} {\bf 107,} 5405--5410 (2010).
%
\bibitem{hopfield_74}
JJ Hopfield,  Kinetic proofreading:  A new mechanism for reducing errors in biosynthetic processes requiring high specificity.  {\em Proc Natl Acad Sci (USA)} {\bf 71,} 4135--4139 (1974).  
%
\bibitem{ninio_75}
J Ninio, Kinetic amplification of enzyme discrimination. {\em Biochimie} {\bf 57,} 587--595 (1975). 
%
\bibitem{hopfield_80}
JJ Hopfield,   The energy relay: A proofreading scheme based on dynamic cooperativity and
lacking all characteristic symptoms of kinetic proofreading in DNA replication and protein
synthesis. {\em Proc Natl Acad Sci (USA)} {\bf 77,} 5248--5252 (1980).
%
\bibitem{bryngelson+wolynes_87}
JD Bryngelson and PG Wolynes, Spin glasses and the statistical mechanics of protein folding.  {\em Proc Natl Acad Sci (USA)} {\bf 84,} 7524--7528 (1987).
%
\bibitem{leopold+al_92}
PE Leopold, M Montal, and JN Onuchic, Protein folding funnels: A kinetic approach to the
sequence--structure relationship.  {\em Proc Natl Acad Sci (USA)} {\bf 89,} 8721--8725 (1992).
%
\bibitem{onuchic+al_95}
JN Onuchic, PG Wolynes, Z Luthey--Schulten, and ND Socci,  Toward an outline of the topography of a realistic protein--folding funnel. {\em Proc Natl Acad Sci (USA)} {\bf 92,} 3626--3630 (1995).
%
\bibitem{desai+al_07}
MM Desai, DS Fisher, and AW Murray, The speed of evolution and maintenance of variation in asexual populations.  {\em Curr Biol} {\bf 17,} 385--394 (2007).
%
\bibitem{neher+al_10}
RA Neher, BI Shraiman, and DS Fisher, Rate of adaptation in large sexual populations. {\em Genetics} {\bf 184,} 467--481 (2010).
%
\bibitem{hallatschek_11}
O Hallatschek, The noisy edge of traveling waves. {\em Proc Natl Acad Sci (USA)} {\bf  108,} 1783--1787 (2011).
%
\bibitem{neher+shraiman_11}
RA Neher and BI Shraiman, Statistical genetics and evolution of quantitative traits.  {\em Rev Mod Phys} {\bf 83,}  1283--1300 (2011).
%
\bibitem{fisher_12}
DS Fisher, Asexual evolution waves:  fluctuations and universality.  arXiv:1210.6295v1 [q--bio.PE] (2012).
%
\bibitem{eguiluz+al_00}
VM Egu\'iluz, M Ospeck, Y Choe, AJ Hudspeth, and MO Magnasco, Essential nonlinearities in hearing.  {\em Phys Rev Lett} {\bf 84,} 5232--5235 (2000).
%
\bibitem{calamet+al_00}
S Calamet, T Duke, F J\"ulicher, and J Prost,  Auditory sensitivity provided by self--tuned critical oscillations of hair cells. {\em Proc Natl Acad Sci (USA)} {\bf 97,} 3183--3187 (2000).
%
\bibitem{magnasco_03}
MO Magnasco,  A traveling wave over a Hopf bifurcation shapes the cochlear tuning curve.
{\em Phys Rev Lett} {\bf 90,} 058101 (2003).
%
\bibitem{vos+al_94}
MH Vos, MR Jones, CN Hunter, J Breton, JC Lambry, and JL Martin, Coherent nuclear dynamics at room temperature in bacterial reaction centers.
 {\em Proc Natl Acad Sci (USA)} {\bf 91,} 12701--12705 (1994).
%
\bibitem{engel+al_07}
GS Engel, TR Calhoun, EL Read, T--K Ahn, T Man\v{c}al, Y--C Cheng,
RE Blankenship, and GR Fleming,  Evidence for wavelike energy transfer through quantum coherence in photosynthetic systems. {\em Nature} {\bf 446,} 782--786 (2007).
%
\bibitem{sturtevant_13}
AH Sturtevant,   The linear arrangement of six sex--linked factors in {\em Drosophila}, as shown by their mode of association. {\em J Exp Zool} {\bf 14,} 43--59 (1913).
%
\end{thebibliography}
\end{document}